\documentclass[runningheads]{llncs}

\usepackage{lineno}
\usepackage{makecell}
\usepackage{multirow}
\usepackage{float}
\usepackage{amssymb}
\usepackage{amsmath, bm}
\usepackage{bm}
\usepackage{algorithm}
\usepackage{algorithmic}
\usepackage{enumerate}
\usepackage{cite}
\usepackage{color}
\usepackage{adjustbox}
\usepackage{blindtext}
\usepackage{inputenc}

\newtheorem{myDef}{Definition}
\usepackage{threeparttable}
\usepackage{booktabs}

\usepackage[hidelinks]{hyperref} 
\hypersetup{
colorlinks=true,
linkcolor=blue,
citecolor=blue
}

\usepackage{graphicx}
\usepackage{tikz,xcolor,hyperref}
\definecolor{lime}{HTML}{A6CE39}
\DeclareRobustCommand{\orcidicon}{%
    \begin{tikzpicture}
    \draw[lime, fill=lime] (0,0) 
    circle [radius=0.16] 
    node[white] {{\fontfamily{qag}\selectfont \tiny ID}};    \draw[white, fill=white] (-0.0625,0.095) 
    circle [radius=0.007];    \end{tikzpicture}
    \hspace{-2mm}}
\foreach \x in {A, ..., Z}{%
    \expandafter\xdef\csname orcid\x\endcsname{\noexpand\href{https://orcid.org/\csname orcidauthor\x\endcsname}{\noexpand\orcidicon}}
    }

\usepackage{marvosym}

\begin{document}



\title{Delving into Cryptanalytic Extraction of \\ PReLU Neural Networks}

\index{Chen, Yi}
\index{Dong, Xiaoyang}
\index{Ma, Ruijie}
\index{Shen, Yantian}
\index{Wang, Anyu}
\index{Yu, Hongbo}
\index{Wang, Xiaoyun}

\titlerunning{Delving into Cryptanalytic Extraction of PReLU Neural Networks}
%




\author{
Yi Chen\inst{1}\orcidA{} \and
Xiaoyang Dong\inst{2,6}\orcidB{} \and
Ruijie Ma\inst{3}\orcidC{} \and
Yantian Shen\inst{3,6}\orcidD{} \and
Anyu~Wang\inst{1,4,5,6}\orcidE{} \and
Hongbo Yu\inst{3,4,6}\orcidG{} \and
Xiaoyun Wang\inst{1,4,5,6,7}\textsuperscript{(\Letter)}\orcidF{}
}

\authorrunning{Y. Chen et al.}


\institute{
Institute for Advanced Study, 
Tsinghua University, Beijing, China, \\
\email{\{chenyi2023, anyuwang, xiaoyunwang\}@tsinghua.edu.cn}
 \\ \and	
Institute for Network Sciences and Cyberspace,
Tsinghua University, Beijing, China,  
\email{xiaoyangdong@tsinghua.edu.cn} \\ \and	
Department of Computer Science and Technology, 
Tsinghua University, Beijing, China,
\email{\{marj21, shenyt22\}@mails.tsinghua.edu.cn}, 
\email{yuhongbo@tsinghua.edu.cn}  \\ \and	
State Key Laboratory of Cryptography and Digital Economy Security, 
Tsinghua University, Beijing, China	\\ \and 
School of Cryptologic Science and Engineering, Shandong University, Jinan, China  \\ \and 
Zhongguancun Laboratory, Beijing, China  \\ \and 
Shandong Institute of Blockchain, Shandong, China	
}

\maketitle              
\begin{abstract}
The machine learning problem of model extraction 
was first introduced in 1991 and 
gained prominence as a cryptanalytic challenge starting with Crypto 2020.  
For over three decades, research in this field has primarily 
focused on ReLU-based neural networks.  
In this work, we take the first step towards the 
cryptanalytic extraction of PReLU neural networks, 
which employ more complex nonlinear activation functions than their ReLU counterparts.

We propose a raw output-based parameter recovery attack for PReLU networks 
and extend it to more restrictive scenarios where 
only the top-m probability scores are accessible.
Our attacks are rigorously evaluated through end-to-end experiments 
on diverse PReLU neural networks, 
including models trained on the MNIST dataset.   
To the best of our knowledge, 
this is the first practical demonstration of PReLU neural network extraction 
across three distinct attack scenarios.

\end{abstract}
\setcounter{footnote}{0}


\section{Introduction}
\label{sec:introduction}

\subsection{Cryptanalytic Extraction of Neural Networks }
The machine learning problem of recovering neural network model parameters
(also known as model extraction)
was initially introduced by Baum in 1991~\cite{DBLP:journals/tnn/Baum91}.
At Crypto 2020, Carlini et al. argued that this machine learning problem 
is a cryptanalytic problem as follows,
and should be studied as such~\cite{DBLP:conf/crypto/CarliniJM20}.

\subsubsection{Problem Statement. }
The victim model (i.e., the neural network, 
denoted by $f_{\theta}$ where $\theta$ is the model parameters) 
is made available as an Oracle.
The adversary can generate any input $X \in \mathcal{X}$ to query the Oracle
and receive the feedback $\zeta$, where $\mathcal{X}$ is the input space of $f_{\theta}$. 
The adversary's target is recovering the model parameters 
by collecting some pairs $(X, \zeta)$.
The model parameters are usually 64-bit floating-point numbers.

Consider that there may be some network isomorphisms 
for the victim model, e.g., 
permutation for fully connected neural 
networks~\cite{DBLP:conf/icml/RolnickK20}.
Therefore, in practice, the adversary aims to recover an 
\emph{equivalent neural network}
~\cite{
DBLP:conf/uss/JagielskiCBKP20,
DBLP:conf/icml/RolnickK20,
DBLP:conf/crypto/CarliniJM20,
DBLP:conf/asiacrypt/ChenDGSWW24,
DBLP:conf/eurocrypt/CarliniCHRS25
}.

\paragraph{\textbf{\textup{Equivalent Neural Networks. }}}
Consider two neural networks $f_{\theta}$ and $f_{\widehat{\theta}}$ 
with different model parameters.
If $f_{\theta} (X) = f_{\widehat{\theta}} (X)$ for all $X \in \mathcal{X}$,
the two neural networks are isomorphic~\cite{DBLP:conf/icml/RolnickK20},
also called functionally equivalent~\cite{DBLP:conf/crypto/CarliniJM20}.

Formally, the target of the cryptanalytic extraction is 
to recover a set of model parameters $\widehat{\theta}$, 
such that $f_{\widehat{\theta}} (X) = f_{\theta} (X)$ for all $X \in \mathcal{X}$, 
i.e., recovering an equivalent neural network. 
Two factors determine the difficulty of achieving the target of cryptanalysis extraction: 
the victim model (concretely, the model architecture and activation functions),
and the \emph{attack scenario}~\cite{DBLP:conf/uss/JagielskiCBKP20}.


\paragraph{\textbf{\textup{Attack Scenarios. }}}
According to the common taxonomy in~\cite{DBLP:conf/uss/JagielskiCBKP20},
for the victim model deployed as a black-box classifier,
there are five attack scenarios defined by
the feedback provided by the Oracle for each input query:
(1) the most likely class label,
(2) the most likely class label along with its probability score,
(3) the top-$m$ labels and their probability scores,
(4) all labels and their probability scores,
(5) the raw output $f_{\theta} (X)$.
Among these feedbacks, 
the raw output leaks the most information, 
and the most likely class label leaks the least~\cite{DBLP:conf/uss/JagielskiCBKP20}.


\subsection{Questions To Be Explored in This Paper}
\label{subsec:questions}

Although the machine learning problem of recovering model parameters 
has been studied for over $30$ years,  the development in this field is relatively slow.
At a high level, there are two common limitations in previous research. 

The first limitation is that most previous research
(including the state-of-the-art work in this field~\cite{DBLP:conf/crypto/CarliniJM20,
DBLP:conf/eurocrypt/CanalesMartinezCHRSS24,
DBLP:conf/asiacrypt/ChenDGSWW24,
DBLP:conf/eurocrypt/CarliniCHRS25
})
has only studied the fully connected neural networks 
based on the ReLU activation function
(shortly called ReLU neural networks in this paper)~\cite{DBLP:conf/iclr/DanielyG23,
DBLP:conf/nips/FoersterMSH24,
DBLP:conf/uss/JagielskiCBKP20,
DBLP:conf/icml/MartinelliSGB24,
DBLP:conf/fat/MilliSDH19,
DBLP:conf/wpes/Reith0T19,
DBLP:conf/uss/TramerZJRR16,
DBLP:conf/crypto/CarliniJM20,
DBLP:conf/eurocrypt/CanalesMartinezCHRSS24,
DBLP:conf/asiacrypt/ChenDGSWW24,
DBLP:conf/eurocrypt/CarliniCHRS25
}.
However, numerous new model architectures 
and activation functions~\cite{DBLP:journals/ijon/DubeySC22}
have emerged in recent decades.


The Parametric Rectified Linear Unit (PReLU) 
activation function~\cite{DBLP:conf/iccv/HeZRS15}
is a typical representative.
As of September 2025, Reference~\cite{DBLP:conf/iccv/HeZRS15} 
has been cited more than 10,000 times
\footnote{Refer to \url{https://ieeexplore.ieee.org/document/7410480}.},
demonstrating the influence of the PReLU activation function.
Compared with the ReLU activation function, 
it does not block the negative input 
and introduces an extra learnable parameter, namely, the slope.
PReLU often works better than ReLU in very deep neural networks,
especially in tasks where negative inputs might carry 
useful information~\cite{DBLP:conf/iccv/HeZRS15, DBLP:journals/ijon/DubeySC22}.

To the best of our knowledge, 
no public research has explored the feasibility of recovering the model parameters 
of any PReLU activation-based neural network.
Therefore, this paper studies the cryptanalytic extraction 
of the PReLU activation function-based
fully connected neural networks (hereafter referred to as PReLU neural networks).

Since the raw output leaks more information than the other four types of feedback, 
we start with the attack scenario where the raw output is accessible.
Hence, the first question to be explored is that
\begin{equation}
\begin{array}{c}
\text{\emph{Is it possible to recover the model parameters of a PReLU neural network}} \\
\text{\emph{given access to its raw output?}} 
\end{array} \nonumber
\end{equation}
Consider that there have been some raw output-based 
model parameter recovery attacks 
(e.g., the state-of-the-art attacks~\cite{DBLP:conf/crypto/CarliniJM20, DBLP:conf/eurocrypt/CanalesMartinezCHRSS24})
designed for ReLU neural networks.
We are curious about the impact of PReLU activation on these attacks,
and wonder whether the core principles of previous attacks
are still useful for PReLU neural networks.

In the state-of-the-art raw output-based attacks~\cite{DBLP:conf/crypto/CarliniJM20, 
DBLP:conf/eurocrypt/CanalesMartinezCHRSS24} against ReLU neural networks,
there are two core tasks,
i.e., \emph{neuron weight recovery} and \emph{neuron sign recovery}.
At CRYPTO 2020,
Carlini et al. proposed a \emph{differential attack} 
for the neuron weight recovery~\cite{DBLP:conf/crypto/CarliniJM20} .
As for the neuron sign recovery, 
Carlini et al. proposed a \emph{preimage-based technique}
that applies to contractive neural networks~\cite{DBLP:conf/crypto/CarliniJM20}.
For expansive neural networks,
the neuron sign is recovered by a brute-force guessing method, 
which requires exponential time complexity~\cite{DBLP:conf/crypto/CarliniJM20}.
At EUROCRYPT 2024,
Canales Martinez et al. proposed a \emph{neuron wiggle technique} 
to recover neuron signs~\cite{DBLP:conf/eurocrypt/CanalesMartinezCHRSS24},
which only requires polynomial time complexity for some expansive neural networks.

%

Thus, the first question to be explored includes three subquestions:
\begin{itemize}
\item
Does the core principle of the neuron weight recovery method 
(i.e., the differential attack)
in~\cite{DBLP:conf/crypto/CarliniJM20} apply to PReLU neural networks?

\item
Do the core principles of the neuron sign recovery methods 
(i.e., the preimage-based technique and the neuron wiggle technique)
in~\cite{DBLP:conf/crypto/CarliniJM20, 
DBLP:conf/eurocrypt/CanalesMartinezCHRSS24} apply to PReLU neural networks?

\item
How do we recover the slope of each PReLU activation function?
\end{itemize}
The first two subquestions focus on
the impact of PReLU activation on previous attacks.
If the core principles of the previous attacks are still applicable, 
we design the attack for PReLU neural networks by adopting the core principles. 
Otherwise, we develop attacks by 
finding new weaknesses in PReLU neural networks.

The second common limitation in previous research is that
the scenario where the feedback is the (top-$m$) scores has not been studied.
Most previous research focuses on the attack scenario where
the feedback is the raw output~\cite{
DBLP:conf/iclr/DanielyG23,
DBLP:conf/nips/FoersterMSH24,
DBLP:conf/uss/JagielskiCBKP20,
DBLP:conf/icml/MartinelliSGB24,
DBLP:conf/fat/MilliSDH19,
DBLP:conf/wpes/Reith0T19,
DBLP:conf/uss/TramerZJRR16,
DBLP:conf/crypto/CarliniJM20,
DBLP:conf/eurocrypt/CanalesMartinezCHRSS24
}.
At ASIACRYPT 2024 and EUROCRYPT 2025,
Chen et al. and Carlini et al., respectively,
have studied the attack scenario where the feedback is
the most likely class label~\cite{
DBLP:conf/asiacrypt/ChenDGSWW24,
DBLP:conf/eurocrypt/CarliniCHRS25
}. 

Therefore, the second question to be explored in this paper is
\begin{equation}
\begin{array}{c}
\text{\emph{Is it possible to recover 
the model parameters of a PReLU neural network}} \\
\text{\emph{given access to the (top-$m$) probability scores?}} 
\end{array} \nonumber
\end{equation}
Once the feasibility of 
recovering model parameters based on the raw output is verified,
we try a new idea, i.e., apply the raw output-based model 
parameter recovery attack to the scenario 
where the feedback is the (top-$m$) probability scores.
The focus is on exploring the feasibility of 
transformation between these attack scenarios.

\subsection{Our Contributions}
The two questions posed in Section~\ref{subsec:questions}
have been addressed in this paper.
We not only demonstrate the feasibility of recovering the model parameters 
of PReLU neural networks under three attack scenarios 
but also elucidate the influence of PReLU activation on prior attacks.
Concretely, we have made contributions in the following four aspects.

\subsubsection{The Influence of PReLU Activation on Previous Attacks. }
We demonstrate that PReLU activation 
has an extremely negative influence on previous attacks 
(Section~\ref{sec:influence_of_prelu}).
Concretely, since PReLU activation does not block negative inputs,
the preimage-based neuron sign recovery method
proposed in~\cite{DBLP:conf/crypto/CarliniJM20} 
does not apply to PReLU neural networks.
Besides, PReLU activation weakens the effectiveness of
the neuron wiggle technique
proposed in~\cite{DBLP:conf/eurocrypt/CanalesMartinezCHRSS24}.
If the neuron wiggle technique is applied to recover the neuron signs of 
PReLU neural networks, 
the time complexity of the end-to-end attack tends to become exponential,
particularly for PReLU neural networks with high-dimensional hidden layers.

\subsubsection{Differential Attack for the Neuron Weight Recovery. }
Following the idea of exploiting the second partial directional derivative  
in~\cite{DBLP:conf/crypto/CarliniJM20},
we propose a differential attack to 
recover the neuron weights of PReLU neural networks 
(Section~\ref{sec:signature_recovery_with_raw_output}),
which demonstrates that the core principle of the differential attack
in~\cite{DBLP:conf/crypto/CarliniJM20} applies to PReLU neural networks.

More importantly, 
we find a common limitation
among differential-based neuron weight recovery attacks 
(including the one in~\cite{DBLP:conf/crypto/CarliniJM20}),
i.e., not applying to highly expansive neural networks,
which has not been reported before.
This finding has unearthed an important open question for future studies.

\subsubsection{Novel Neuron Sign and Slope Recovery Algorithms. }
In this paper, one of the core contributions is that we propose two new methods to recover the 
neuron sign and slope simultaneously (Section~\ref{sec:slope_recovery_raw_output}), 
considering that the existing neuron sign recovery techniques 
lose their power due to PReLU activations.
The two new methods reveal different weaknesses of PReLU neural networks.

Concretely, the first method shows that the equivalent affine transformations 
of the PReLU neural network in two adjacent linear regions leak the slope
of the PReLU activation.
The second one shows that,
if we adjust the PReLU neural network
by a \emph{neuron splitting technique} newly proposed in this paper,
the second partial directional derivative of 
the PReLU neural network at critical points also leaks the slope.

\subsubsection{Top-m Scores-based Model Parameter Recovery Attack. }
The newly proposed differential attack, 
as well as the methods for the neuron sign and slope recovery,
are both based on the raw output of PReLU neural networks.
Based on a newly proposed technique, namely, the \emph{neuron fusion technique},
we successfully apply the raw output-based model parameter recovery attack
to the scenario where the feedback is the (top-$m$) scores 
(Section~\ref{sec:attack_with_scores}).
Moreover, by proposing a minor improvement on 
the binary search-based method in~\cite{DBLP:conf/crypto/CarliniJM20} 
for finding critical points,
we show that the choice of $m$ 
has no significant influence on the proposed model parameter recovery attack.  

We have performed adequate practical end-to-end experiments 
(Section~\ref{sec:practical_experiments})
on various PReLU neural networks trained on random data 
and MNIST (a widely used benchmarking dataset in visual deep learning),
which verifies the effectiveness of the proposed model parameter recovery attacks.
Our attacks require polynomial query and time complexity, 
see Appendix~\ref{appendix:attack_complexity} for details.
All the experiments presented in this paper can be completed using a single core of a modern computer within a practical time.
The code is accessible in \url{https://github.com/AI-Lab-Y/Extracting_PReLU_NN}.

\section{Preliminaries}
\label{sec:preliminaries}

\subsection{Basic Definitions and Notations}


\begin{myDef} 
[\textbf{\textup{$n$-Deep Neural Network}}]
\label{def:k-deep-nn}
A \textup{$n$-deep neural network} $f_{\theta}$ is a function 
parameterized by $\theta$ that takes inputs from an input space $\mathcal{X}$
and returns values in an output space $\mathcal{Y}$.
The function $f \colon \mathcal{X} \rightarrow \mathcal{Y}$
is composed of alternating linear layers $f_i$ and non-linear
activation layers $\sigma_i$:
\begin{equation}
f = f_{n+1} \circ \sigma_n \circ \cdots \circ \sigma_2 \circ f_2 \circ \sigma_1 \circ f_1 .
\end{equation}
\end{myDef}


\begin{myDef} 
[\textbf{\textup{Fully Connected Layer}}]
The $k$-th \textup{fully connected layer} of a neural network is
a function $f_k \colon \mathbb{R}^{d_{k-1}} \rightarrow \mathbb{R}^{d_k}$
given by the affine transformation
\begin{equation}
f_k (X) = A^{k} X + B^{k}.
\end{equation}
where $A^{k} \in \mathbb{R}^{d_k \times d_{k-1}}$ 
is a $d_k \times d_{k-1}$ \textup{weight} matrix,
$B^{k} \in \mathbb{R}^{d_k}$ is a $d_k$-dimensional \textup{bias} vector,
and $X = [x_1, x_2, \cdots, x_{d_k - 1}]^{\top}$ is the input vector.
\end{myDef}

Denote
\begin{eqnarray}
A^k = 
\begin{bmatrix}
a_{1,1}^{k}		& 	a_{1,2}^{k}		&	\cdots	&	a_{1, d_{k-1}}^{k}		\\
a_{2,1}^{k}		& 	a_{2,2}^{k}		&	\cdots	&	a_{2, d_{k-1}}^{k}		\\
\vdots                 &	\vdots                &                         &       \vdots                        \\
a_{d_k, 1}^{k}	&	a_{d_k, 2}^{k}	&	\cdots        &	a_{d_k, d_{k-1}}^{k}	\\
\end{bmatrix},  \,\,\,
B^k = 
\begin{bmatrix}
b_1^{k} \\
b_2^{k} \\
\vdots  \\
b_{d_k}^k \\
\end{bmatrix}
\end{eqnarray}
where $a_{i, j}^{k}, b_i^k$ 
$\left( 1 \leqslant i \leqslant d_k, 1 \leqslant j \leqslant d_{k-1} \right)$
are floating numbers.
In the rest of the paper,
we denote by $A_i^k$ the $i$-rh row of $A^k$,
and $A_{:,i}^k$ the $i$-th column of $A^k$.

\begin{myDef} 
[\textbf{\textup{Activation Layer}}]
The $k$-th \textup{activation layer} $\sigma_k$ of a neural network is defined by
$d_k$ non-linear activation functions $\sigma_{k, i}$ for $i \in \{1, \cdots, d_k\}$.
\end{myDef}

In this paper, we focus on the PReLU~\cite{DBLP:conf/iccv/HeZRS15} 
activation function
\begin{equation}
\sigma _{k,i} (x) = \left\{
\begin{array}{l}
x, \,\,\,\,\,\,\,\,\,\,\,\,\,\, \mathrm{if} \,\, x \geqslant 0 ,\\
s_i^k \times x, \, \mathrm{if} \,\, x < 0 ,\\
\end{array}
\right.
\end{equation}
where $s_i^k > 0$ is the slope determined by the training of neural networks. 
Denote by $Y^k = \left[ y_1^k, y_2^k, \cdots, y_{d_k}^k \right]^{\top}$
the preactivation input of $\sigma _k$ in layer $k$.
The $\sigma _k$ layer can be expressed as a matrix $I^k$ multiplied by $Y^k$
\begin{eqnarray}
\sigma _k (Y^k) = I^k Y^k,  \,\,\,
I^k = 
\begin{bmatrix}
\tau _1^{k}		& 	0		&	\cdots	&	0		\\
0			& 	\tau _2^{k}		&	\cdots	&	0		\\
\vdots                 &	\vdots                &                         &       \vdots                        \\
0			&	0		&	\cdots        &	\tau_{d_k}^{k}	\\
\end{bmatrix}, \,\,\,
\tau _i^k = \left\{
\begin{array}{l}
1, \,\,\,\, \mathrm{if} \,\, y_i^k \geqslant 0 ,\\
\\
s_i^k , \, \mathrm{if} \,\, y_i^k < 0 .\\
\end{array}
\right.
\end{eqnarray}

\begin{myDef} 
[\textbf{\textup{Neuron}}]
A \textup{neuron} is a function determined by the corresponding
weight matrix, bias vector, and activation function.
Formally, the $i$-th neuron of layer $k$ is the function $\eta_i^k$ given by
\begin{equation}
\eta_i^k (\sigma_{k-1} (Y^{k-1})) = 
\sigma _{k,i} \left( A_i^{k} \sigma_{k-1} (Y^{k-1}) + b_i^{k} \right),
\end{equation} 
where $\sigma_{k-1} (Y^{k-1})$ is the output of the $(k-1)$-th activation layer. 
There are $d_i$ neurons in layer $i$.
\end{myDef}

\begin{myDef} 
[\textbf{\textup{Model Architecture}}]
The \textup{architecture} of a fully connected  neural network 
captures the structure of $f_{\theta}$: (a) the number of layers, 
(b) the dimension $d_k$ of each layer $k \in \{0, \cdots , n+1\}$.
We say that $d_0$ and $d_{n+1}$ are the input and output dimensions
of the neural network.
\end{myDef}

In the following, the $n$-deep neural network $f_{\theta}$ is denoted by 
'$d_0$-$d_1$-$\cdots$-$d_{n+1}$', 
where $d_j$ is the dimension in layer $j$.

\begin{myDef} 
[\textbf{\textup{Model Parameters}}]
The \textup{parameters} $\theta$ of a $n$-deep neural network $f_{\theta}$ 
are the concrete assignments
to the weights $A^{k}$, biases $B^{k}$ for $k \in \{1, \cdots, n+1\}$,
and slopes $S^k$ for $k \in \{1, \cdots, n \}$.
Here, $S^k = [s_1^k, s_2^k, \cdots, s_{d_k}^k].$
\end{myDef}

\begin{myDef} 	
[\textbf{\textup{Critical Point}}]	\label{def:critical_point}
A \textup{critical point} is an input $X \in \mathbb{R}^{d_0}$ 
that makes the input of an activation function of a neuron $0$.
\end{myDef}

For example,
if $X$ makes $A_i^{k} \sigma_{k-1} (Y^{k-1}) + b_i^{k} = 0$,
then $X$ is a critical point corresponding to
the $i$-th neuron in layer $k$.
In Sections~\ref{sec:signature_recovery_with_raw_output}
and~\ref{sec:slope_recovery_raw_output},
for the convenience of introducing our work,
we also say that $Y^{k-1}$ or $\sigma_{k-1} (Y^{k-1})$ is a critical point
corresponding to the $i$-th neuron in layer $k$.

\begin{myDef} [\textbf{\textup{Neuron State}}]
Let $\mathcal{V}(\eta; X)$ denote the value that neuron $\eta$ takes
with $X \in \mathbb{R}^{d_0}$ before applying its corresponding activation function $\sigma$. 
If $\mathcal{V}(\eta; X) > 0$ (respectively, $\mathcal{V}(\eta; X) < 0$),
the \textup{neuron state} of $\eta$ is \textup{positive} (respectively, \textup{negative} ).
If $\mathcal{V}(\eta; X) = 0$,
the neuron state is \textup{critical}.
\end{myDef}


\subsection{Adversarial Goals and Assumptions}
\label{subsec:goal_and_assumptions}

As introduced in Section~\ref{sec:introduction},
there are two parties in a model parameter recovery attack:
the Oracle and the adversary.
By generating any input $X$ to query the Oracle,
the adversary receives a feedback $\zeta$. 
Based on a certain number of pairs of $X$ and $\zeta$, 
the adversary recovers a set of parameters $\widehat{\theta}$ 
with the goal that $f_{\widehat{\theta}}(x)$ 
behaves as same as possible to $f_{\theta}(x)$ in terms of some metrics.
In this paper,
we use the $\widehat{\_}$ symbol to indicate a recovered parameter.

\paragraph{\textup{\textbf{Assumptions. }}}
We make the following assumptions about Oracle and the capabilities of the attacker:
\begin{itemize}
\item 
\textbf{Architecture knowledge.  } 
We require knowledge of the model architecture of the neural network.

\item 
\textbf{Chosen inputs. } 
We can feed arbitrary inputs from $\mathcal{X} = \mathbb{R}^{d_0}$.

\item
\textbf{Precise computations. }
$f_{\theta}$ is specified and evaluated 
using 64-bit floating-point arithmetic.


\item
\textbf{PReLU Activations. }
All activation functions $\sigma_{i, j}$ are the PReLU function 
with a slope $0 < s_j^i < 1$.
\end{itemize}

Compared with previous work 
in~\cite{DBLP:conf/uss/JagielskiCBKP20,
DBLP:conf/icml/RolnickK20,
DBLP:conf/crypto/CarliniJM20,
DBLP:conf/eurocrypt/CanalesMartinezCHRSS24,
DBLP:conf/asiacrypt/ChenDGSWW24,
DBLP:conf/eurocrypt/CarliniCHRS25
},
we replace the ReLU activation function with the PReLU activation function.
According to~\cite{DBLP:conf/iccv/HeZRS15}, almost all the slopes are between 0 and 1.
As a first step to evaluate the risk of model extraction of PReLU neural networks,
we focus on the case of $0 < s_i^k < 1$ in this paper.
Removing this restriction would be an important research direction in the future.


\section{Influence of PReLU Activation on Previous Attacks}
\label{sec:influence_of_prelu}

As introduced in Section~\ref{subsec:questions}, 
in the case where the feedback is the raw output,
the state-of-the-art attacks against ReLU neural networks are proposed 
in~\cite{DBLP:conf/crypto/CarliniJM20, DBLP:conf/eurocrypt/CanalesMartinezCHRSS24}.
At a high level, the two main steps 
in~\cite{DBLP:conf/crypto/CarliniJM20, DBLP:conf/eurocrypt/CanalesMartinezCHRSS24}
and their targets are as follows:
\begin{itemize}
\item \textbf{Neuron weight recovery: } Denote by $\widehat{A}_i^k$ 
the final recovered weight of the $i$-th neuron in layer $k$.
The \emph{neuron weight recovery} aims to recover 
$\widetilde{A}_i^k \in \{\widehat{A}_i^k,  -\widehat{A}_i^k\}$.

\item \textbf{Neuron sign recovery: } After the neuron weight recovery,
the next step is the \emph{neuron sign recovery}, i.e.,
checking whether $\widehat{A}_i^k = \widetilde{A}_i^k$ or
$\widehat{A}_i^k = - \widetilde{A}_i^k$.
\end{itemize}


For the neuron weight recovery,
Carlini et al. in~\cite{DBLP:conf/crypto/CarliniJM20} 
proposed a \emph{differential attack}.
In Section~\ref{sec:signature_recovery_with_raw_output}, 
we will show how to extend it to PReLU neural networks
and present an analysis of the implications of PReLU activation.

This section introduces the negative impact of PReLU activation on
the previous neuron sign recovery techniques,
including the \emph{preimage-based technique} 
in~\cite{DBLP:conf/crypto/CarliniJM20}
and the \emph{neuron wiggle technique}  
in~\cite{DBLP:conf/eurocrypt/CanalesMartinezCHRSS24}.

%

\subsection{Influence on the Preimage-based Technique in~\cite{DBLP:conf/crypto/CarliniJM20}}
\label{subsec:impact_on_sign_recovery}

\subsubsection{Preimage-based Technique~\cite{DBLP:conf/crypto/CarliniJM20}. }
Consider the $i$-th neuron $\eta_i^k$ in layer $k$.

The preimage-based technique proposed in~\cite{DBLP:conf/crypto/CarliniJM20} is as follows.
Assume that $X$ is a critical point corresponding to the neuron $\eta_i^k$,
and $h^k = [h_1^k, h_2^k, \cdots, h_{d_k}^k]$ is the preactivation input caused by $X$ to
the $k$-th activation layer, where $h_i^k = 0$.
The adversary computes the preimage $\widetilde{X}$ corresponding to
a new preactivation input $\widetilde{h}^k = [\widetilde{h}_1^k, 
\widetilde{h}_2^k, \cdots, \widetilde{h}_{d_k}^k]$ to
the $k$-th activation layer, where
\begin{equation}
\widetilde{h}_j^k = \left\{ 
\begin{array}{l}
h_j^k,   \,\, \textup{if} \,\, j \ne i  ,\\
\varDelta ,  \,\,\, \textup{if} \,\, j = i . \\
\end{array}
\right.
\end{equation}
When the neuron sign is $1$,
(i.e., $\widehat{A}_i^k = \widetilde{A}_i^k$),
one has $f_{\theta} (X) = f_{\theta} (\widetilde{X})$ 
if $\varDelta < 0$.
When the neuron sign is $-1$,
(i.e., $\widehat{A}_i^k = -\widetilde{A}_i^k$),
one has $f_{\theta} (X) = f_{\theta} (\widetilde{X})$ 
if $\varDelta > 0$.

\subsubsection{The Impact of PReLU Activation. }
The above preimage-based technique utilizes the property that 
the negative preactivation input of a neuron
does not influence the raw output of the neural network,
since ReLU activation blocks negative inputs.
PReLU activation does not block the negative input,
thus, the negative input influences the raw output,
making the preimage-based technique inapplicable to PReLU neural networks.

\subsection{Influence on the Neuron Wiggle 
Technique in~\cite{DBLP:conf/eurocrypt/CanalesMartinezCHRSS24} }

\subsubsection{Neuron Wiggle Technique~\cite{DBLP:conf/eurocrypt/CanalesMartinezCHRSS24}. }
The core idea of the neuron wiggle technique is that 
a small perturbation (i.e., the neuron wiggle)
in a carefully chosen direction is expected to 
change the target neuron's output by $\varDelta$, 
while changing all other neurons in the same layer 
by about $\pm \frac{\varDelta}{\sqrt{d}}$,
where $d$ is the dimension of the target layer~\cite{DBLP:conf/eurocrypt/CanalesMartinezCHRSS24}.

More concretely, 
when the preactivation input of the target neuron is positive,
the change $v_{+}$ of the output in the target layer is 
\begin{equation}
v_{+} = \left( \pm \frac{\varDelta}{\sqrt{d}}, \pm \frac{\varDelta}{\sqrt{d}}, 
		   \cdots, \varDelta, \cdots,  
		   \pm \frac{\varDelta}{\sqrt{d}}, \pm \frac{\varDelta}{\sqrt{d}} \right)
\end{equation}
When the preactivation input of the target neuron is negative,
the corresponding change $v_{-}$ is
\begin{equation}
v_{-} = \left( \pm \frac{\varDelta}{\sqrt{d}}, \pm \frac{\varDelta}{\sqrt{d}}, 
		   \cdots, 0, \cdots,  
		   \pm \frac{\varDelta}{\sqrt{d}}, \pm \frac{\varDelta}{\sqrt{d}} \right)
\end{equation}
According to the analysis in~\cite{DBLP:conf/eurocrypt/CanalesMartinezCHRSS24},
the change caused by $v_{+}$ in the raw output of the ReLU neural network
is more likely to be larger than that caused by $v_{-}$, 
since $\Vert v_{+} \Vert_2 > \Vert v_{-} \Vert_2$.
By randomly sampling some critical points 
and finding a neuron wiggle for each critical point,
one can recover the neuron sign by 
a counting method~\cite{DBLP:conf/eurocrypt/CanalesMartinezCHRSS24}.

In~\cite{DBLP:conf/nips/FoersterMSH24}, 
Foerster et al. conduct an in-depth analysis of the neuron wiggle technique 
and present two important findings.
First, it is common that the neuron wiggle technique 
fails to recover some neuron signs, 
which is also verified by the experiments 
in~\cite{DBLP:conf/eurocrypt/CanalesMartinezCHRSS24}. 
Second, due to various reasons 
(e.g., some neurons being too \textquoteleft close\textquoteright to each other), 
it is hard to promote the success rate of the neuron wiggle technique 
by increasing the number of randomly sampled critical points.

However, in a practical end-to-end model parameter recovery attack,
if we can not correct these sign recovery errors in layer $i$, 
the end-to-end attack will fail since the recovery of deeper layers is
sensitive to errors~\cite{
DBLP:conf/crypto/CarliniJM20,
DBLP:conf/asiacrypt/ChenDGSWW24
}.
Thus, Foerster et al. in~\cite{DBLP:conf/nips/FoersterMSH24}
suggest correcting the sign recovery errors
(i.e., the neuron signs that are recovered incorrectly by the neuron wiggle technique)
by the brute-force guessing method 
introduced in~\cite{DBLP:conf/crypto/CarliniJM20}.

If the number of neuron signs recovered incorrectly is small,
the above suggested method works, 
and the resulting time complexity may still be practical.
Otherwise, the end-to-end attack becomes impractical.

\subsubsection{The Impact of PReLU Activation. }
If we apply the neuron wiggle technique to PReLU neural networks,
due to the slope of the PReLU activation function,
the two changes $v_{+}$ and $v_{-}$ are expected to be
\begin{equation}
\begin{aligned}
v_{+} &= \left( \pm \frac{\tau_1 \varDelta}{\sqrt{d}}, \pm \frac{ \tau_2 \varDelta}{\sqrt{d}}, 
		   \cdots, \varDelta, \cdots,  
		   \pm \frac{\tau_{d-1} \varDelta}{\sqrt{d}}, \pm \frac{\tau_d \varDelta}{\sqrt{d}} \right)  ,	\\
v_{-} &= \left( \pm \frac{\tau_1  \varDelta}{\sqrt{d}}, \pm \frac{\tau_2 \varDelta}{\sqrt{d}}, 
		   \cdots, s_i \varDelta, \cdots,  
		   \pm \frac{\tau_{d-1} \varDelta}{\sqrt{d}}, \pm \frac{\tau_d \varDelta}{\sqrt{d}} \right)  ,	\\
\end{aligned}
\end{equation}
where $s_i$ is the slope of the PReLU activation in the target neuron,
and $\tau_j \in \{1, s_j\}$ for $j \ne i$.
These slopes weaken the advantage of $v_{+}$, 
making the sign of the target neuron hard to recover.
Particularly, when $s_i$ approaches $1$, 
the difference $\Vert v_{+} \Vert_2 - \Vert v_{-} \Vert_2$
is even likely to approach $0$.


As a result, the number of neuron signs recovered incorrectly 
is likely to increase significantly, i.e., the failure rate increases,
resulting in the risk of the time complexity being exponential, 
and the end-to-end attack becoming impractical.

\subsubsection{Supporting Experiment. }
To verify the impact of PReLU activation, 
we have performed a support experiment as follows.
Given a PReLU neural network,
we perform the neuron sign recovery attack following the attack setting
\footnote{
In~\cite{DBLP:conf/eurocrypt/CanalesMartinezCHRSS24},
Shamir et al. assume that all the parameters (excluding the neuron signs in layer $i$) 
of the first $i$ layers have been recovered without any errors,
and the number of critical points for recovering a neuron sign is $200$.
}
used in~\cite{DBLP:conf/eurocrypt/CanalesMartinezCHRSS24},
and focus on the success rate of the neuron sign recovery, 
i.e., how many neuron signs are recovered correctly (and incorrectly) in layer $i$.

\begin{table}[!htbp]
\centering
\renewcommand\arraystretch{1.0}
\caption{The results of neuron sign recovery attacks on PReLU neural networks.}
\label{tab:support_experiment}
\begin{tabular}{c|c|c|c|c}
\hline
\makecell{Architecture \\ (training dataset)}	&	
\makecell{Layer 1 \\ $\checkmark$ / $\times$ \\ (Neuron wiggle)}		&	
\makecell{Layer 2 \\ $\checkmark$ / $\times$ \\ (Neuron wiggle)}		&
\makecell{Layer 1 \\ $\checkmark$ / $\times$ \\ (Our method)}		&
\makecell{Layer 2 \\ $\checkmark$ / $\times$ \\ (Our method)}		\\
\hline
\makecell{196-50-50-$1^{!}$ \\ (random data)}	&	
48 / 2	&	35 / 15	&	50 / 0	&	50 / 0	\\
\hline
\makecell{196-100-100-$1^{!}$ \\ (random data)}	&	
86 / 14	&	72 / 28	&	100 / 0	&	100 / 0	\\
\hline
\makecell{196-50-50-$1^{\star}$ \\ (random data)}	&	
44 / 6	&	33 / 17	&	50 / 0	&	50 / 0	\\
\hline
\makecell{196-100-100-$1^{\star}$ \\ (random data)}	&	
82 / 18	&	69 / 31	&	100 / 0	&	100 / 0	\\
\hline
\makecell{200-150-150-$1^{\star}$ \\ (random data)}	&	
110 / 40	&	108 / 42	&	150 / 0	&	150 / 0	\\
\hline
\makecell{250-200-200-$1^{\star}$ \\ (random data)}	&	
151 / 49	&	148 / 52	&	200 / 0	&	200 / 0	\\
\hline
\makecell{784-200-200-$1^{\star}$ \\ (MNIST, `7' vs `3')}	&	
173 / 27	&	168 / 32	&	200 / 0	&	200 / 0	\\	
\hline
\makecell{784-210-210-$1^{\star}$ \\ (MNIST, `9' vs `2')}&	
181 / 29	&	162 / 48	&	210 / 0	&	210 / 0	\\
\hline
\multicolumn{5}{l}{
Our method: the proposed method in Section~\ref{subsec:joint_recovery}.  
}		\\
\multicolumn{5}{l}{
$!$: all the neuron slopes of the victim model satisfy the condition $0 < s_j^i < 1.0$. 
}		\\
\multicolumn{5}{l}{
$\star$: all the neuron slopes of the victim model satisfy the condition $0.9 < s_j^i < 1.0$. 
}		\\
\multicolumn{5}{l}{
$\checkmark$ / $\times$: the number of neuron signs recovered correctly (incorrectly)
}		\\
\multicolumn{5}{l}{
MNIST, `$c_1$' vs `$c_2$': the training dataset is MNIST; the target is to classify $c_1$ and $c_2$.
}		\\
\end{tabular}
\end{table}

The support experiment is performed on several PReLU neural networks
trained on random data and a real image dataset 
(MNIST~\cite{DBLP:journals/pieee/LeCunBBH98}, 
a typical benchmarking dataset in visual deep learning).
Table~\ref{tab:support_experiment} summarizes the experiment results.
As a comparison, the same experiments are performed
on our neuron sign recovery method
\footnote{
To recover a neuron sign, 
our method requires an average of only $\epsilon \rm{log} _2 {d_i}$ 
(much less than 200) critical points, where $\epsilon$ is a small constant,
refer to Section~\ref{subsec:joint_recovery} and the analysis of the attack complexity
in Appendix~\ref{appendix:attack_complexity}.
}
introduced in Section~\ref{subsec:joint_recovery}.
From these results, we obtain the following important findings:
\begin{itemize}
\item
First, the neuron wiggle technique fails to recover some neuron signs,
and the failure rate is not low,
even if we do not make any constraints on the neuron slopes,
refer to rows 2 and 3.
Moreover, with the dimension (the number of neurons in layer $i$) increasing,
the failure rate also increases.

\item
Second, in the case where all the slopes in layer $i$ exceed $0.9$,
the failure rate of the neuron wiggle technique is much higher,
refer to the remaining PReLU neural networks in rows $4, \cdots, 9$.
\end{itemize}


We have conducted additional experiments on 
two representative models (250-200-200-1 and 784-210-210-1).  
Specifically, we increase the number of critical points
(for recovering each neuron sign) to 2000 and 4000, respectively.  
However, the experimental results demonstrate 
no improvement in the success rate,
i.e.,  the neuron signs recovered incorrectly
in Table~\ref{tab:support_experiment}
are still incorrect.

The experimental results verify the negative influence of PReLU activations on
the neuron wiggle technique.
By adopting the suggestion given in~\cite{DBLP:conf/nips/FoersterMSH24},
i.e., correct the sign recovery errors by the brute-force guessing-based method 
introduced by Carlini et al. in~\cite{DBLP:conf/crypto/CarliniJM20},
one may correct all the sign recovery errors
\footnote{It is unclear whether the brute-force guessing method 
in~\cite{DBLP:conf/crypto/CarliniJM20} applies to PReLU neural networks.
Even if it is applicable, the time complexity is still too high.},
but the time complexity would be exponential
and the attack would be impractical.


\section{Overview of Our Attacks}
\label{sec:attack_overview}

This paper first develops model parameter recovery attacks leveraging the raw output, 
and subsequently explore their application in scenarios 
where only the (top-$m$) probability scores are accessible.
This section presents an overview of the raw output-based 
model parameter recovery attack, 
for better grasping the principles of our attacks.

Besides, Appendix~\ref{sec:properties_of_prelu}
summarizes the equivalence of PReLU neural networks 
and the linear region property of PReLU neural networks, 
which are the bases of our attacks. 
We recommend that readers refer to Appendix~\ref{sec:properties_of_prelu} 
if there are any puzzles in the following reading.

\subsubsection{Main Steps of Our Model Parameter Recovery Attack.}
Our attack recovers the model parameters layer by layer.
In layer $k \in \{1, \cdots, n\}$,
the parameter recovery of the $i$-th neuron is composed of two steps:
\begin{itemize}
\item \textbf{Neuron weight recovery: } recover the weights and bias up to
some unknown multiplicative factor, 
i.e., obtaining $\widetilde{A}_i^k = c_i^k A_i^k$. 

\item \textbf{Neuron sign and slope recovery: } 
recover the slope $s_i^k$ and the neuron sign (i.e., the sign of the factor $c_i^k$),
such that we obtain $\widehat{A}_i^k = \left| c_i^k \right| A_i^k$. 
\end{itemize}
For the last layer (i.e., layer $n+1$),
it only involves the weight and bias recovery. 


In this paper, 
based on two different findings, 
we propose two methods to 
recover the sign and slope of a neuron simultaneously.
The first method (see Section~\ref{subsec:independent_recovery}) 
is based on the finding that the equivalent affine functions 
of the neural network in two adjacent linear regions leak the slope.
The second method (see Section~\ref{subsec:joint_recovery}) 
is based on the finding that
the second partial directional derivative of 
the PReLU neural network at critical points also leaks the slope, 
if we make some adjustments to the neural network.

When using the first method to recover slopes,
there are no effective ways to refine the precision of the recovered slopes so far.
The adverse effect is that the recovery error in the slopes of 
the first activation layer tends to be relatively significant, 
eventually leading to a large recovery error (or even a failed recovery) 
for the model parameters of deeper layers.

Based on a newly proposed technique, namely,
the neuron splitting technique,
the second method transforms the recovery of the neuron sign and slope
into the neuron weight recovery,
which makes it possible to refine the precision of the recovered slopes
by refining the precision of neuron weights.
Our experimental results also show that
the second method makes the model parameter recovery attack performs better.

\begin{figure}[!htb]
\centering
\includegraphics[width=0.8\textwidth]{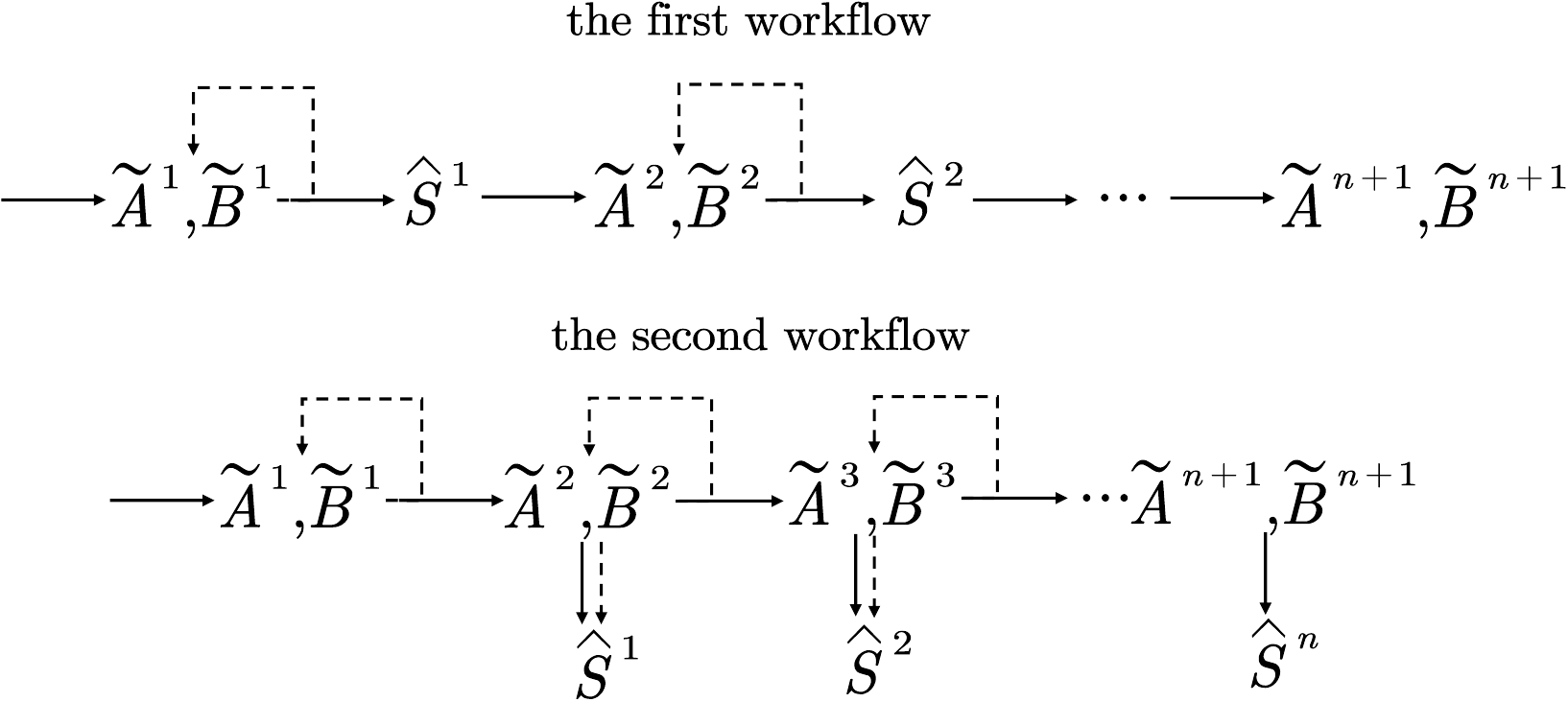}
\caption{
Two workflows of the raw output-based model parameter recovery attack. 
Each solid arrow represents recovering the parameters, 
and each dotted arrow represents refining the precision of recovered parameters. 
}
\label{fig:two_workflows}
\end{figure}

\paragraph{\textbf{\textup{Two Different Workflows.  }}}
The two slope recovery methods 
lead to two different workflows of the model parameter recovery attack.
To help understand our attacks, 
we describe the two workflows as depicted in Fig.~\ref{fig:two_workflows}.

When using the first method in Section~\ref{subsec:independent_recovery}
(respectively, the second method in Section~\ref{subsec:joint_recovery}), 
the entire model parameter recovery attack follows the first workflow 
(respectively, the second workflow). 
The precision of recovered slopes in each layer is not refined in Workflow 1
but is refined in Workflow 2 by refining the weights of the next layer.
Note that there is an exception.
When the victim model is a $1$-deep PReLU neural network,
the recovered slopes $\widehat{S}^1$ are not further refined,
since the recovered weights $\widehat{A}^2$ of the last layer will not be refined.



\section{Extend Differential Attack on ReLU Neural Networks to PReLU Neural Networks}
\label{sec:signature_recovery_with_raw_output}

%
%

This section first shows how the neuron weights
are recovered by the differential attack designed for PReLU neural networks,
then analyze the common limitation of differential attacks
(including the one in~\cite{DBLP:conf/crypto/CarliniJM20}),
as well as the impact of PReLU activation.
For the convenience of introducing the attack,
we assume that the output dimension is $d_{n+1} = 1$, 
i.e., $f_{\theta} (X) \in \mathbb{R}$. 
Note that this assumption is not fundamental,
and our attack applies to the case of $d_{n+1} > 1$.
Besides, some auxiliary algorithms (i.e., finding and filtering critical points,
refining the precision of recovered neuron weights)
are presented in Appendix~\ref{appendix:auxiliary_algos}.

In this paper, we compute the second partial directional derivatives using finite differences. 
Other methods that can estimate the derivatives accurately are also applicable.

\subsection{Recover Hidden Layers}	
\label{subsec:recover_hidden_layer}

Given an input $X$, 
the raw output $f_{\theta} (X)$ is computed as
\begin{equation}	\label{eq:n_deep_nn_initial}
f_{\theta}(X) = A^{n+1} \cdots 
                           \left( A^2 I^1\left( A^1 X + B^1 \right) + B^2 \right) 
			\cdots + B^{n+1} .
\end{equation}

\subsubsection{Core Idea. }
Without loss of generality,
consider the recovery of the first neuron $\eta_1^i$ in layer $i \in \{2, \cdots, n\}$,
and assume that the first $i-1$ layers have been recovered.
Note that the method in this section applies to layer $1$.

Let $X$ be a critical point corresponding to the neuron $\eta_1^i$,
and denote by $Z \in \mathbb{R}^{d_{i-1}}$
the resulting input of the $i$-th fully connected layer
(i.e., the output of the $(i-1)$-th activation layer).
Then, we have $A_1^i Z + b_1^i = 0$.

Next, let us focus on the unrecovered part of the neural network,
and skip the first $i-1$ layers.
Consider two adjacent linear regions $\mathcal{R}_{+}$ and $\mathcal{R}_{-}$, 
with $Z$ located at the boundary between them.
For $\forall Z_{+} \in \mathcal{R}_{+}$, the neuron state of $\eta_1^i$ is positive.
For $\forall Z_{-} \in \mathcal{R}_{-}$, the neuron state of $\eta_1^i$ is negative.
The neuron states of the remaining neurons (except for $\eta_1^i$) are
the same for $Z$, $Z_{+}$, and $Z_{-}$.

Let $Z$ move in two opposite directions
\footnote{This is achieved by letting $X$ move in two opposite directions.} 
(denoted by $H \in \mathbb{R}^{d_{i-1}}$ and $-H$) 
with a small stride $\epsilon$,
such that $Z + \epsilon H$ and $Z - \epsilon H$ are still in the two adjacent linear regions
\footnote{This condition is easy to achieve. 
If the stride $\epsilon$ is not small enough, 
the subsequent attack will give a random incorrect result.
Such an incorrect result behaves like noise (i.e., only occurring once) 
and will be abandoned.
},
i.e., $Z + \epsilon H \in \mathcal{R}_{+}$ and $Z - \epsilon H \in \mathcal{R}_{-}$
(or $Z + \epsilon H \in \mathcal{R}_{-}, Z - \epsilon H \in \mathcal{R}_{+}$).
Query the neural network with $Z$, $Z + \epsilon H$,
and $Z - \epsilon H$, we get
\begin{equation}
\begin{aligned}
f_{\theta} (Z) &= 
G^{i+1} I^i (A^i  Z + B^i)  + U^{i+1} ,  \\
f_{\theta} (Z + \epsilon H) &= 
G^{i+1} I_{+}^i (A^i  (Z + \epsilon H) + B^i)  + U^{i+1} ,  \\
f_{\theta} (Z - \epsilon H) &= 
G^{i+1} I_{-}^i (A^i  (Z - \epsilon H) + B^i)  + U^{i+1} ,  \\
\end{aligned}
\end{equation}
where
\begin{equation}
I^i = 
\begin{bmatrix}
1 	     & 		0 	 & 	\cdots      &	0 	\\
0 	     & \tau_2^i	 & 	\cdots      &  	0		\\
\vdots    &   \vdots      &   		       &   	\vdots       \\
0	     &		0	&	\cdots      &	\tau_{d_i}^i	\\
\end{bmatrix}  ,   \,\,\,
I_{+}^i = 
\begin{bmatrix}
\tau _{1+}^i 	     & 		0 	 & 	\cdots      &	0 	\\
0 	     & \tau_2^i	 & 	\cdots      &  	0		\\
\vdots    &   \vdots      &   		       &   	\vdots       \\
0	     &		0	&	\cdots      &	\tau_{d_i}^i	\\
\end{bmatrix} ,  \,\,\,
I_{-}^i = 
\begin{bmatrix}
\tau _{1-}^i     & 		0 	 & 	\cdots      &	0 	\\
0 	     & \tau_2^i	 & 	\cdots      &  	0		\\
\vdots    &   \vdots      &   		       &   	\vdots       \\
0	     &		0	&	\cdots      &	\tau_{d_i}^i	\\
\end{bmatrix} ,
\end{equation}
and
\begin{equation}
\left( \tau_{1+}^i,  \tau_{1-}^i \right) = \left\{ \begin{array}{l}
\left( 1, s_1^i \right) ,  \,\, \mathrm{if} \,\, A_1^i H > 0,	\\
	\\
\left( s_1^i, 1 \right),   \,\, \mathrm{if} \,\, A_1^i H < 0.
\end{array} \right.
\end{equation}
Note that $G^{i+1} = [g_1^{i+1}, \cdots, g_{d_i}^{i+1}] \in \mathbb{R}^{d_i}$ 
and $U^{i+1} \in \mathbb{R}$ are determined by the critical point $X$
(more concretely, determined by all the \emph{neuron states} in layer $i+1$ to $n$).

Then,
the second partial directional derivative $\delta$ of $f_{\theta}$ at $Z$ is
\begin{small}
\begin{equation}	\label{eq:second_order_diff}
\delta = 
\frac{
f_{\theta} \left( Z + \epsilon H \right)  + 
f_{\theta} \left( Z - \epsilon H \right) 
- 2 f_{\theta} \left( Z \right)
}{
\epsilon
}	
=
\left\{  \begin{array}{l}
g_{1}^{i+1} \left( 1 - s_1^i \right) A_1^i  H,
\,\, \mathrm{if} \,\, A_1^i  H >0 ,\\
	\\
g_{1}^{i+1} \left( s_1^i - 1 \right) A_1^i H ,
\,\, \mathrm{if} \,\, A_1^i  H <0 .\\
\end{array}
\right.
\end{equation}
\end{small}

The relation in Eq.~\eqref{eq:second_order_diff}  inspires us to recover
the neuron weight by the following method.
Choose $d_{i-1}$ directions $H_j$ and
compute the value $\delta _j$ as shown in Eq.~\eqref{eq:second_order_diff},
then build a system of linear equations based on 
$\delta_j$ and $H_j$, solve it for the neuron weight.

However, we do not know the sign of $A_1^i  H_j $,
which prevents us from building a system of linear equations.
To tackle this problem,
we first assume that $A_1^i H_1 > 0$.
Then, based on this assumption,
we recover the sign of $A_1^i H_j$ where $j \in \{2, \cdots, d_{i-1}\}$.
If the assumption is wrong,
we flip the sign of $A_1^i H_j$.

\subsubsection{Recover the sign of $A_1^i H_j$. }
Next, we show how to recover the sign of $A_1^i H_j$,
under the assumption that $A_1^i H_1 > 0$.

Query $f_{\theta}$ with $Z + \epsilon (H_1 + H_j)$
and $Z - \epsilon (H_1 + H_j)$, and let
\begin{equation}
\delta_{1+j} = \frac{
f_{\theta} \left( Z + \epsilon \left( H_1 + H_j \right) \right) +
f_{\theta} \left( Z - \epsilon \left( H_1 + H_j \right) \right) -
2 f_{\theta} \left( Z \right)
}{
\epsilon
} .	\nonumber
\end{equation}
The absolute value of $\delta_{1+j}$ is 
\begin{equation}
\left|	\delta_{1+j} \right| = 
\left|
g_{1}^{i+1} (1 - s_1^i) A_1^i (H_1 + H_j)
\right|		\nonumber
\end{equation}

Next, consider two absolute values 
$\left|  \delta_1 + \delta_j \right|$,
$\left|  \delta_1 - \delta_j \right|$,
we have
\begin{equation}
\left.  \begin{array}{r}
\left|  \delta_1 + \delta_j \right| = 
\left| g_{1}^{i+1} (1 - s_1^i)  A_1^i (H_1 + H_j)  \right|  \\
\left|  \delta_1 - \delta_j \right|  =
\left| g_{1}^{i+1} (1 - s_1^i) A_1^i  (H_1 - H_j)  \right|
\end{array}
\right\} , \,\, \mathrm{if} \,\, (A_1^i H_1) (A_1^i H_j) > 0 ,	\nonumber
\end{equation}
\begin{equation}
\left.  \begin{array}{r}
\left|  \delta_1 + \delta_j \right| = 
\left| g_{1}^{i+1} (1 - s_1^i)  A_1^i (H_1 - H_j)  \right|  \\
\left|  \delta_1 - \delta_j \right|  =
\left| g_{1}^{i+1} (1 - s_1^i) A_1^i  (H_1 + H_j)  \right|
\end{array}
\right\} , \,\, \mathrm{if} \,\, (A_1^i H_1) (A_1^i H_j) < 0 .	\nonumber
\end{equation}
Therefore, we check whether $(A_1^i H_1) (A_1^i H_j) > 0$
by the following rules 
\begin{equation}
\left\{
\begin{array}{l}
\left| \left| \delta_{1+j} \right| - \left| \delta_1 + \delta_j \right| \right|
<
\left| \left| \delta_{1+j} \right| - \left| \delta_1 - \delta_j \right| \right|, 
\,\, \mathrm{if} \,\,  (A_1^i H_1) (A_1^i H_j) > 0  ,		\\
	\\
\left| \left| \delta_{1+j} \right| - \left| \delta_1 + \delta_j \right| \right|
>
\left| \left| \delta_{1+j} \right| - \left| \delta_1 - \delta_j \right| \right|, 
\,\, \mathrm{if} \,\,  (A_1^i H_1) (A_1^i H_j) < 0	.	\\
\end{array}
\right.
\end{equation}

\subsubsection{Recover the Neuron Weight $\widehat{A}_1^i$. }
After determining the sign of $A_1^i H_j$, 
we build a system of linear equations
and solve for the vector $\widetilde{A}_1^i$ such that 
$\widetilde{A}_1^i \cdot H_j = \alpha _j$,
where $\alpha _j = \delta _j$ or $\alpha _j = - \delta _j$
is determined by the sign of $A_1^i H_j$.

Note that the weight vector $\widetilde{A}_1^i$
is obtained based on the assumption of
$A_1^i H_1 > 0$.
This assumption may be wrong,
then the recovered weight vector should be 
$\widehat{A}_1^i = - \widetilde{A}_1^i$.
In other words, there is \emph{one bit of sign information} 
(i.e., the neuron sign) to be recovered.
After obtaining the weight vector $\widetilde{A}_1^i$,
we compute
$\widetilde{b}_1^{i} = - \widetilde{A}_1^i Z$.
And the final recovered bias is 
$\widehat{b}_1^i = \pm \widetilde{b}_1^i$.

\subsection{Recover The Last Layer}
\label{subsec:recover_last_layer_output}
Next, consider the last layer.
Denote by $Z$ the output of the $n$-th activation layer,
one has $f_{\theta} (Z) = A^{n+1}  Z + B^{n+1}$.
To obtain the weight vector $\widehat{A}^{n+1} \in \mathbb{R}^{d_n}$ 
and  bias $\widehat{B}^{n+1} \in \mathbb{R}$, 
we randomly sample $d_n + 1$ inputs $Z$, 
collect the outputs $f_{\theta} (Z)$,
build a system of linear equations, 
and solve it for $\widehat{A}^{n+1}$ and $\widehat{B}^{n+1}$.


\subsection{Limitation and The Impact of PReLU Activation}
The attack proposed in Section~\ref{subsec:recover_hidden_layer} illustrates that 
the fundamental principle of the differential attack~\cite{DBLP:conf/crypto/CarliniJM20}, 
originally developed for ReLU neural networks, 
is equally applicable to PReLU neural networks. 
However, our attack shares the same limitation 
as the differential attack described in~\cite{DBLP:conf/crypto/CarliniJM20}, 
with the PReLU activation function further exacerbating these constraints.

\subsubsection{Common Limitation. }
The common limitation is that the differential attacks, 
including the one proposed in~\cite{DBLP:conf/crypto/CarliniJM20} for ReLU neural networks
and our attack in this section, do not apply to highly expansive neural networks
\footnote{This limitation is not reported in~\cite{DBLP:conf/crypto/CarliniJM20} 
and newly found in this paper.
}.

Consider the recovery of layer $i$ where $i \geqslant 2$,
and suppose that 
\emph{
$d_{i-2}$ is the smallest one 
among the dimensions of the first $i-2$ layers}. 
The core idea of our differential attack and the one in~\cite{DBLP:conf/crypto/CarliniJM20} 
is to build a system of linear equations 
and solve it for a weight vector $W$. 
Denote by $\widetilde{d}_{i-1}$ (respectively, $\widetilde{d}_{i-2}$) 
the number of unknown weight parameters in the system of linear equations
(respectively, the number of degrees of freedom at the output of layer $i-2$).
As a result, our differential attack and 
the one in~\cite{DBLP:conf/crypto/CarliniJM20}
work only if $\widetilde{d}_{i-1} \leqslant \widetilde{d}_{i-2}$. 

For ReLU neural networks,
according to~\cite{DBLP:conf/crypto/CarliniJM20, DBLP:conf/eurocrypt/CanalesMartinezCHRSS24},
we have $\widetilde{d}_{i-1} \approx \frac{1}{2} d_{i-1}$
and $\widetilde{d}_{i-2} \leqslant d_{i-2}$.
Therefore, we find that the differential attack in~\cite{DBLP:conf/crypto/CarliniJM20} 
does not apply to highly expansive neural networks where $d_{i-1} \gg 2 d_{i-2}$,
which has been verified on some ReLu neural networks (e.g., 10-30-30-1),
by running the code in~\cite{DBLP:conf/crypto/CarliniJM20}.

\subsubsection{Impact of PReLU Activation. }
Since PReLU activation does not block negative inputs,
we have $\widetilde{d}_{i-1} = d_{i-1}$ and $\widetilde{d}_{i-2} = d_{i-2}$
\footnote{
The equality may not hold if the weight matrix corresponding to
 the affine transformation of the recovered layers has colinear vectors,
but this happens with low probability.
}
for PReLU neural networks,
which further exacerbates the limitation,
i.e., our attack does not work if $d_{i-1} > d_{i-2}$.

For now, there are no effective methods to break this limitation
for ReLU and PReLU neural networks.
Besides, consider that this limitation does not influence the conclusion that
the core principle of the differential attack in~\cite{DBLP:conf/crypto/CarliniJM20}
applies to PReLU neural networks, we leave it as future research.

%
%
%
%
%

\section{Raw Output-based Sign and Slope Recovery}
\label{sec:slope_recovery_raw_output}

The attacks in Section~\ref{sec:signature_recovery_with_raw_output} 
recover the weights of each neuron up to an unknown non-zero factor, 
i.e., we obtain $c_j^i A_j^i$ where $c_j^i \in \mathbb{R}$.
To satisfy the requirement of equivalent networks,
we further recover $\widehat{A}_j^i = \left| c_j^i \right| A_j^i$,
i.e., recover the neuron sign.
At the same time, we need to recover 
the slope $s_j^i$ of the PReLU activation function.
This section introduces two methods 
to recover the neuron sign $c_j^i$ and slope $s_j^i$ simultaneously.
Again, for convenience, suppose that $f_{\theta} (X) \in \mathbb{R}$.

\subsection{SOE-based Independent Recovery}
\label{subsec:independent_recovery}


When recovering the neuron signs and slopes in layer $i$,
we assume that the first $i-1$ layers have been recovered,
and the weights and biases of layer $i$ are recovered up to a factor.
The first method processes a neuron each time.

\subsubsection{Core Idea. }
First, let us see how the slope is leaked.
Consider the first neuron $\eta_1^i$ in layer $i$.
We regard the union of the $i$-th activation layer
and the continuous $n+1-i$ layers (i.e., layers $i+1, \cdots, n+1$) 
as a piecewise affine function
\begin{equation}	
f_{\theta} (Y) = G^{i+1} I^i Y + U^{i+1}  ,
\end{equation}
where $Y = [y_1, \cdots, y_{d_i}]^{\top} \in \mathbb{R}^{d_{i}}$ is the input 
of the $i$-th activation layer,
$U^{i+1} \in \mathbb{R}$ and 
$G^{i+1} \in \mathbb{R}^{d_i}$
are determined by all the neuron states in layers $i+1$ to $n$.

Denote by $g_{k}^{i+1}$ the $k$-th element of $G^{i+1}$, we have
\begin{equation}	\label{eq:nn_for_sign_recovery}
f_{\theta} (Y) = 
\sum_{j=1}^{d_i}{g_j^{i+1} \tau _j^i y_{j}} + U^{i+1} .
\end{equation}
Note that $\tau _j^i = s_j^i$ if $y_j < 0$, 
and otherwise $\tau _j^i = 1$.

Next, consider two adjacent linear regions 
$\mathcal{R}_{+}$ and $\mathcal{R}_{-}$.
For $\forall Y_{+} \in \mathcal{R}_{+}$, the output of $\eta_1^i$ is positive.
For $\forall Y_{-} \in \mathcal{R}_{-}$, the output of $\eta_1^i$ is negative.
All the other neuron states in the two linear regions are the same.
Then, in the two linear regions, the corresponding affine functions are
\begin{equation}	\label{eq:affine_trans}
\begin{aligned}
f_{\theta}(Y_+) &= g_1^{i+1} y_{1} + 
			       \sum_{j=2}^{d_i}{g_j^{i+1} \tau _j^i y_{j}} + 
			       U^{i+1} = W_{+}^i Y + U^{i+1}  , \\
f_{\theta}(Y_{-}) &= g_1^{i+1} s_1^{i} y_{1} + 
			       \sum_{j=2}^{d_i}{g_j^{i+1} \tau _j^i y_{j}} + 
			       U^{i+1} = W_{-}^i Y + U^{i+1}   .   \\
\end{aligned}
\end{equation}
It is clear that the first coefficients are different and they leak the slope.

\subsubsection{Recover Affine Transformations. }
Suppose that $Y = \left[ y_1, y_2, \cdots,
y_{d_{i}} \right]^{\top}$ is a critical point
corresponding to the first neuron in layer $i$, i.e., $y_1 = 0$.

Randomly sample $d_i + 1$ points $Y + H^j$
in the neighborhood of $Y$
\begin{equation}
Y + H^j = \left[
y_1 + h_{1}^j,  y_2 + h_{2}^j,  \cdots,
y_{d_i} + h_{d_{i}}^j  
\right]^{\top} , \,\, j \in \{1, \cdots, d_i + 1\} .
\end{equation}
We require that $h_{1}^j > 0$ and $H^j$ 
does not change the neuron states in layer $i+1$ to $n$.
This is easy to achieve if $\Vert H^j \Vert_1$ is small enough.
As a result, if the unknown non-zero factor is positive, i.e., $c_1^i > 0$, 
all the $d_i + 1$ points belong to $\mathcal{R}_{+}$.
If $c_1^i < 0$, all the $d_i + 1$ points belong to $\mathcal{R}_{-}$.

From the raw output $f_{\theta} (Y + H^j)$, we build
a system of linear equations 
\begin{equation}	\label{eq:soe_sign_slope}
W_{+}^i \cdot \left( H^j - H^1 \right) = 
f_{\theta} (Y + H^j) - 
f_{\theta} (Y + H^1)
\,\, \mathrm{for} \,\,  j \in \{2, \cdots, d_i + 1\} .
\end{equation}
Solve it for the weight vector 
$W_{+}^i = \left[ w_{1+}^i, w_{2+}^i, \cdots, w_{d_i +}^i \right]$
where 
\begin{equation}
\left\{ \begin{array}{l}
w_{1+}^i = g_{1}^{i+1},  \,\,\,\,\,\,\,\,\,\,\,\, \mathrm{if} \,\, c_1^i > 0	 , \\
\\
w_{1+}^i = - g_{1}^{i+1} s_1^i ,  \,\, \mathrm{if} \,\, c_1^i < 0  .
\end{array}
\right.
\end{equation}
Note that $d_{i-1} \geqslant d_i$ must hold. 

Similarly, randomly sample $d_i + 1$ points $Y + H^j$ where 
\begin{equation}
H^j = \left[ h_{1}^j, h_{2}^j, \cdots, h_{d_{i}}^j \right]^{\top}, \,\,
h_{1}^j < 0, \,\, j \in \{ 1, \cdots, d_i + 1\},
\end{equation}
build the system of linear equations in Eq.~\eqref{eq:soe_sign_slope},
and solve it for the weight vector
$W_{-}^i = \left[ w_{1-}^i, w_{2-}^i, \cdots, w_{d_i -}^i \right]$
where 
\begin{equation}
\left\{ \begin{array}{l}
w_{1-}^i = g_{1}^{i+1} s_1^i,  \,\, \mathrm{if} \,\, c_1^i > 0  ,	\\
\\
w_{1-}^i = - g_{1}^{i+1},  \,\, \mathrm{if} \,\, c_1^i < 0  .
\end{array}
\right.
\end{equation}

\subsubsection{Recover the Neuron Sign and Slope. }
Note that $\left| g_1^{i+1} s_1^i \right| < \left| g_1^{i+1} \right|$,
since $0 < s_1^i < 1$.
As a result, the slope $\widehat{s}_1^i$ and the sign of $c_1^i$ are determined
by the following rules
\begin{equation}
\mathrm{sign} \left( c_1^i \right) = \left\{
\begin{array}{l}
1, \,\, \mathrm{if} \,\, \left| w_{1+}^i \right| > \left| w_{1-}^i \right|  \\
\\
-1, \,\, \mathrm{if} \,\, \left| w_{1+}^i \right| < \left| w_{1-}^i \right|
\end{array}
\right.	,
\,\,\,
\widehat{s}_1^i = \left\{
\begin{array}{l}
\left| 
\frac{w_{1-}^i}{w_{1+}^i}
\right|,	
\,\, \mathrm{if} \,\, \left| w_{1+}^i \right| > \left| w_{1-}^i \right|	\\
\\
\left|
\frac{w_{1+}^i}{w_{1-}^i}
\right|,
\,\, \mathrm{if} \,\, \left| w_{1+}^i \right| < \left| w_{1-}^i \right|
\end{array}
\right.  .
\end{equation}
The recovered slope $\widehat{s}_1^i$ 
is the same as that $s_1^i$ of the victim model 
(refer to Eq.~\eqref{eq:affine_trans}),
which satisfies the requirement posed by the network equivalence.


\subsection{SOE-based Joint Recovery}
\label{subsec:joint_recovery}

The first method in Section~\ref{subsec:independent_recovery}
exploits the affine transformations in two adjacent linear regions.
We find that the second partial directional derivative of 
the PReLU neural network at critical points also leaks the slope,
which is based on a newly proposed technique, namely,
\emph{Neuron Splitting Technique}.

\begin{figure}[htb]
\centering
\includegraphics[width=0.8\textwidth]{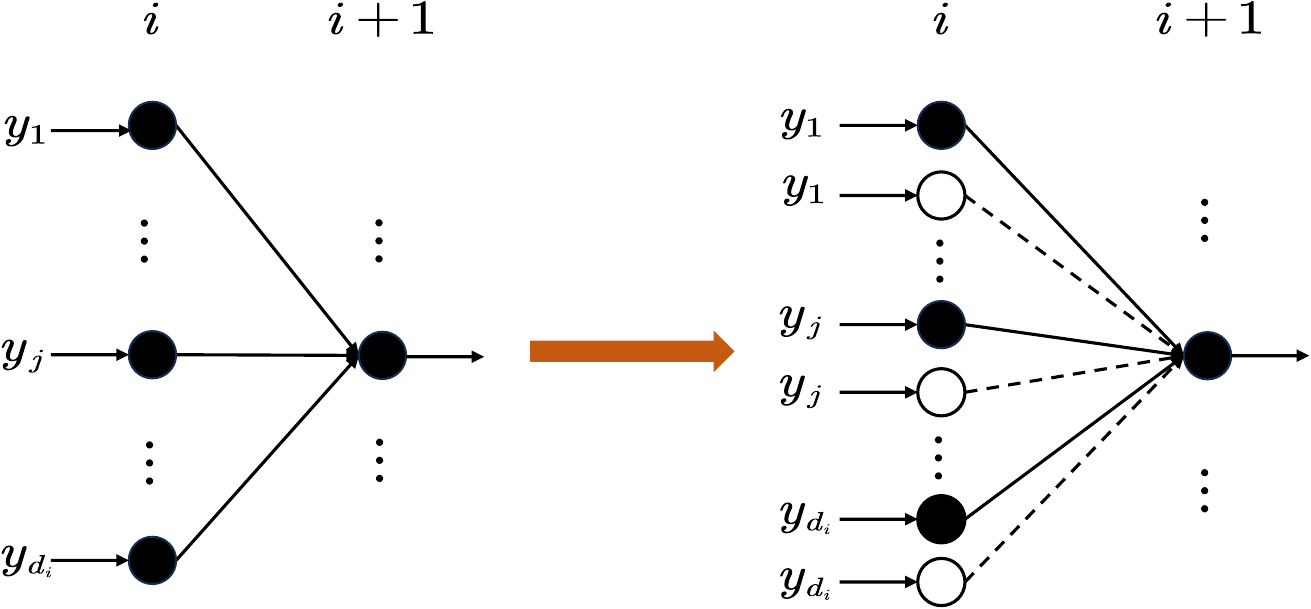}
\caption{
The schematic diagram of the neuron splitting technique.
Expand layer $i$ by adding a twin neuron (the hollow circle) for each neuron.
}
\label{fig:expand_nn}
\end{figure}

\paragraph{\textbf{\textup{Neuron Splitting Technique.  }}}
The neuron splitting technique, as depicted in Fig.~\ref{fig:expand_nn},
expands the target layer (layer $i$)
by making some adjustments to layers $i-1$, $i$, and $i+1$:
\begin{itemize}
\item expand layer $i$ by adding a twin neuron 
(the hollow circle, denoted by $\eta _{j + d_i}^i$) 
for the $j$-th neuron (the solid circle, denoted by $\eta _j^i$).

\item Connect layer $i$ with layer $i+1$. 
Focus on a pair of neurons, e.g., $\eta _j^i$ and $\eta_{j+d_i}^i$.
The weight of the connection between $\eta_{j}^i$ and $\eta_{k}^{i+1}$
is $a_{k, j}^{i+1}$. 
We let the weight of the connection between $\eta_{j + d_i}^i$ 
and $\eta_{k}^{i+1}$ be $s_{j}^i a_{k, j}^{i+1}$. 

\item Connect layer $i-1$ with layer $i$.
Set the same weight vector and bias for each twin neuron
according to the initial neuron,
ensuring the inputs to a pair of neurons are the same. 

\item Change the activation function for each neuron in layer $i$.
For each initial neuron, set the activation function
$\sigma _j^i (y_j) = \max \{y_j, 0\}$.
For each twin neuron, set the activation function
$\sigma _j^i (y_j) = \min \{y_j, 0\}$.
This setting ensures that only one output of the two neurons is non-zero.
\end{itemize}
Now, when $y_j > 0$, the output of the neuron $\eta_j^i$ is $y_j$,
and the raw output of the neural network is not changed.
If $y_j < 0$, the output of the neuron $\eta_{j + d_i}^i$ is $y_j$,
and the raw output is also unchanged since
the weight of the connection between $\eta_{j + d_i}^i$ 
and $\eta_k^{i+1}$ is $s_j^i a_{k, j}^{i+1}$. 

These adjustments lead to the following fact.
If we can recover the new weight vector (see Eq.~\eqref{eq:new_w}) 
of the $k$-th neuron in layer $i+1$,
\begin{equation}	\label{eq:new_w}
\left[ a_{k, 1}^{i+1}, a_{k, 2}^{i+1}, \cdots, a_{k, d_i}^{i+1}, 
         s_1^i a_{k, 1}^{i+1}, s_2^i a_{k, 2}^{i+1}, \cdots, 
         s_{d_i}^i a_{k, d_i}^{i+1} \right]^{\top}  ,
\end{equation}
all the $d_i$ neuron slopes of layer $i$ are also recovered simultaneously. 
All the $d_i$ neuron signs can also be recovered,
which will be introduced later.


\subsubsection{Recover Expanded Weight Vectors. }
To recover the extended weight vector, 
we adjust the workflow for the whole model parameter recovery attack.

When using the first method, 
we first recover $A^i$ and $B^i$ up to a factor, 
subsequently, recover the neuron signs and slopes $S^i$, 
and then proceed to the recovery of the subsequent layer.
Now, the new workflow is as follows.
After recovering $A^i$ and $B^i$ up to a factor,
we adjust the neural network by the neuron splitting technique,
and directly recover layer $i+1$,
i.e., recover the extended weight vector.
Then, recover the neuron signs and slopes $S^i$ in layer $i$.
The new workflow does not change the rationale of weight and bias recovery
proposed in Section~\ref{sec:signature_recovery_with_raw_output},
except for making a little change to the process.

Consider the recovery of the extended weight vector
(denoted by $W = \left[ w_1, \cdots, w_{d_i}, w_{d_i + 1},
 \cdots, w_{2 d_i} \right]^{\top}$)
of the $k$-th neuron in layer $i+1$.
Based on a critical point corresponding to this neuron,
we recover $d_i$ elements of the extended weight vector,
using the attack proposed in 
Section~\ref{sec:signature_recovery_with_raw_output}.
Concretely, for $j \in \{1, \cdots, d_i\}$,
we will obtain the $j$-th element $w_j$ if $y_j > 0$,
or the $(j + d_i)$-th element $w_{d_i + j}$ if $y_j < 0$.
By unifying multiple critical points corresponding to
the same critical points
\footnote{When two different critical points 
corresponding to the same neuron in layer $i+1$, 
the weight vector returned by the attack in 
Section~\ref{sec:signature_recovery_with_raw_output} 
will share common values.
}, 
we recover all the $2 d_i$ elements.

Note that the neuron splitting technique can not overcome the drawback that
the differential attack in Section~\ref{sec:signature_recovery_with_raw_output}
does not apply to expansive PReLU neural networks,
since the number of active neurons (i.e., neurons with non-zero output)
in layer $i$ is always $d_i$ and $d_i > d_{i-1}$ holds for expansive networks.

\subsubsection{Recover the Neuron Sign and Slope. }
Next, let us see how to recover the neuron signs and slopes of layer $i$.
Consider a pair of elements of the extended weight vector $w_j$ and $w_{j + d_i}$.
The neuron sign (i.e., the sign of $c_j^i$) and slope $s_j^i$
of the $j$-th neuron in layer $i$ are as follows
\begin{equation}	\label{eq:two_cases}
\mathrm{sign} \left( c_j^i \right) = \left\{
\begin{array}{l}
1, \,\, \mathrm{if} \,\, \left| w_j \right| > \left| w_{j + d_i} \right|  \\
\\
-1, \,\, \mathrm{if} \,\, \left| w_j \right| < \left| w_{j + d_i} \right|
\end{array}
\right.	,
\,\,\,
\widehat{s}_1^i = \left\{
\begin{array}{l}
\left|
\frac{w_{j + d_i}}{w_j}
\right|,	
\,\, \mathrm{if} \,\, \left| w_{j} \right| > \left| w_{j + d_i} \right|	\\
\\
\left|
\frac{w_{j}^i}{w_{j + d_i}}
\right|,
\,\, \mathrm{if} \,\, \left| w_{j} \right| < \left| w_{j + d_i} \right|
\end{array}
\right.   .
\end{equation}

As for the extracted weight between 
$\eta _j^i$ (the $j$-th neuron in layer $i$) and 
$\eta _k^{i+1}$ (the $k$-th neuron in layer $i+1$),
we let it be $w_j$ if $\left| w_{j} \right| > \left| w_{j + d_i} \right|$,
or $w_{j + d_i}$ if $\left| w_{j} \right| < \left| w_{j + d_i} \right|$.

\subsubsection{Comparison with the First Method in Section~\ref{subsec:independent_recovery}. }
The first method only applies to non-expansive neural networks,
since it requires $d_{i-1} \geqslant d_i$.
The second method transforms the slope recovery into the neuron weight recovery.
Therefore, if there are neuron weight recovery methods for 
expansive PReLU neural networks in the future,
the second method still applies to expansive neural networks.

Besides, the precision refinement algorithm proposed in~\cite{DBLP:conf/crypto/CarliniJM20}
is general and adopted in this paper to refine the precision of recovered neuron weights.
Now, if we use the second method to recover the slopes,
the precision of the recovered slopes is also refined simultaneously.


\section{Scores-based Model Parameter Recovery Attacks}
\label{sec:attack_with_scores}
When the feedback is the scores,
we argue that the model parameters can be recovered 
by transforming the given attack scenario into one 
where the raw output is accessible. 
This section proposes a simple yet effective method to achieve the target.

\subsection{Transform Scores into Raw Output}

\subsubsection{The Case of $d_{n+1} = 1$. }
When the output dimension is $d_{n+1} = 1$,
the raw output $f_{\theta} (X) \in \mathbb{R}$ 
is processed by the Sigmoid function  
to the score $q$
\begin{equation}
q = \frac{1}{ 1 + e^{- f_{\theta} (X)} } .
\end{equation}
In this case, 
we directly compute the raw output $f_{\theta} (X)$ from the score $q$.

\subsubsection{The Case of $d_{n+1} > 1$. }
When the output dimension is $d_{n+1} > 1$,
the scores $Q \in \mathbb{R}^{d_{n+1}}$ are usually obtained
by applying the Softmax function  
to the raw output $f_{\theta} (X) \in \mathbb{R}^{d_{n+1}}$.
Let $Y = f_{\theta} (X) = [y_1, y_2, \cdots, y_{d_{n+1}}]$
and $Q = [q_1, q_2, \cdots, q_{d_{n+1}}]$, then
\begin{equation}	\label{eq:softmax}
q_i = \frac{e^{y_i}}{\sum_{j=1}^{d_{n+1}}{e^{y_j}}} .
\end{equation}

In the case of $d_{n+1} > 1$, 
we can not directly recover the raw output from the scores. 
However, the difference between $y_{i_1}$ and $y_{i_2}$ is
\begin{equation}	\label{eq:new_raw_output}
y_{i_1} - y_{i_2} = \ln \left( \frac{q_{i_1}}{q_{i_2}} \right),
\end{equation}
where $i_1, i_2 \in \{1, \cdots, d_{n+1}\}$.
Therefore, the raw output-based model parameter
recovery attack is still applicable in this attack scenario,
which is achieved by a \emph{neuron fusion technique}
and the fact that the last layer is linear.

\subsubsection{Neuron Fusion Technique. }
\begin{figure}[!htb]
\centering
\includegraphics[width=0.9\textwidth]{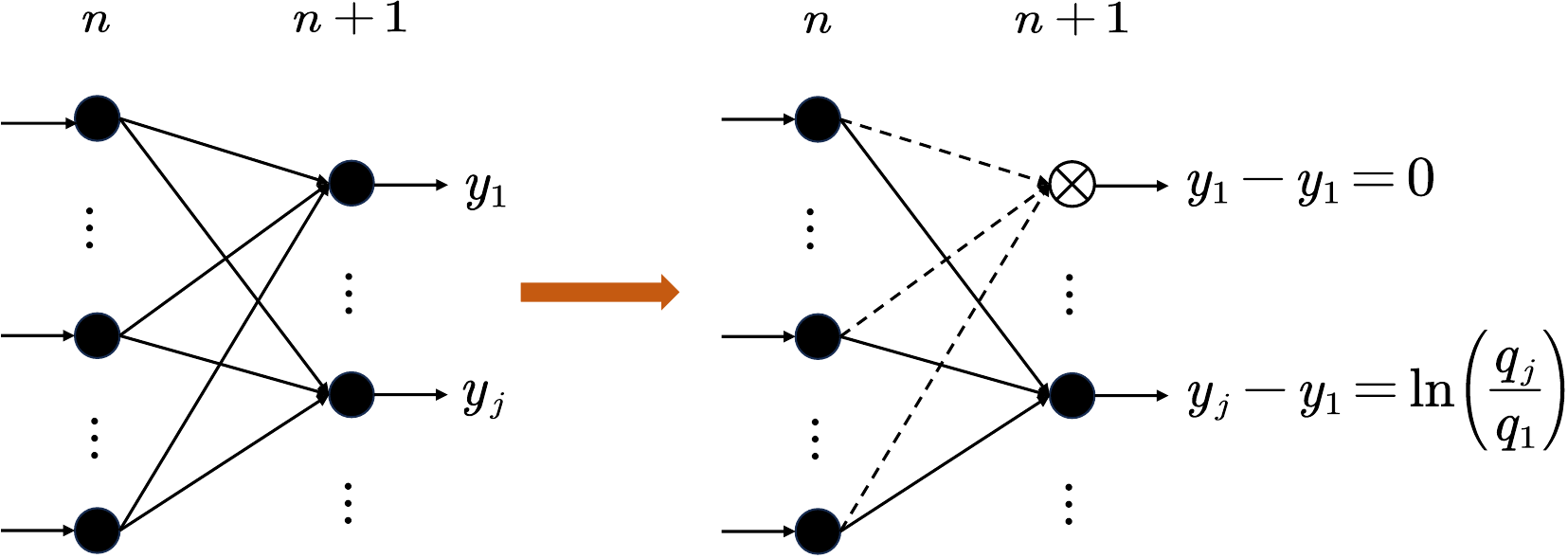}
\caption{
The schematic diagram of the neuron fusion technique.
}
\label{fig:scores_based_attack}
\end{figure}

The idea of the neuron fusion technique is
illustrated in Fig.~\ref{fig:scores_based_attack}.

We make the following adjustment to layer $n+1$ (the last layer):
\begin{itemize}
\item Select a neuron, e.g., the $1$-st neuron in the last layer,
change the weight vector and bias of the $j$-th neuron in the last layer
\begin{equation}
A_j^{n+1} \leftarrow A_j^{n+1} - A_1^{n+1},  \,\,
b_j^{n+1} \leftarrow b_j^{n+1} - b_1^{n+1}.
\end{equation}
\end{itemize}
Now, the raw output of the $j$-th neuron 
in the last layer is $y_j - y_1 = \ln \left( \frac{q_j}{q_1} \right)$.
As a result, the raw output-based model parameter recovery attack is applicable.
The final recovered neural network 
is equivalent to the adjusted neural network.
This adjustment does not affect the recovery of previous layers.


\subsection{The Influence of Top-$m$ Scores}
When the adversary only receives the top-$m$ class labels 
and their probability scores, i.e., the feedback is the third type,
one can still apply the raw output-based
model parameter recovery attack based on the neuron fusion technique.
However, this feedback poses a new problem.

\subsubsection{The Negative Influence on Finding Critical Points. }
Denote by $f_{\theta} (X) = [y_1, y_2, \cdots, y_{d_{n+1}}]$,
$Q = [q_1, q_2, \cdots, q_{d_{n+1}}]$, respectively,
the raw output of the victim model and the probability scores.

In~\cite{DBLP:conf/crypto/CarliniJM20,
DBLP:conf/eurocrypt/CanalesMartinezCHRSS24},
to find a critical point using a binary search-based method, 
the adversary samples two starting points $X_1$ and $X_2$,
and then search critical points using an element $y_j$ of the raw outputs.
 
Let $\mathcal{L}_1$ and $\mathcal{L}_2$ 
denote the sets of top-$m$ labels returned for $X_1$ and $X_2$, respectively. 
If the number of elements of the intersection 
$\mathcal{L}_1 \cap \mathcal{L}_2 $ is less than 2, 
transforming the scores $q_{j_1}$ for $j_1 \in \mathcal{L}_1$ 
(and similarly, $q_{j_2}$ for $j_2 \in \mathcal{L}_2$) 
into the raw output (as defined in Eq.~\eqref{eq:new_raw_output}) 
of the adjusted neural network  
results in a lack of common elements to identify critical points.

\begin{figure}[!htb]
\centering
\includegraphics[width=0.5\textwidth]{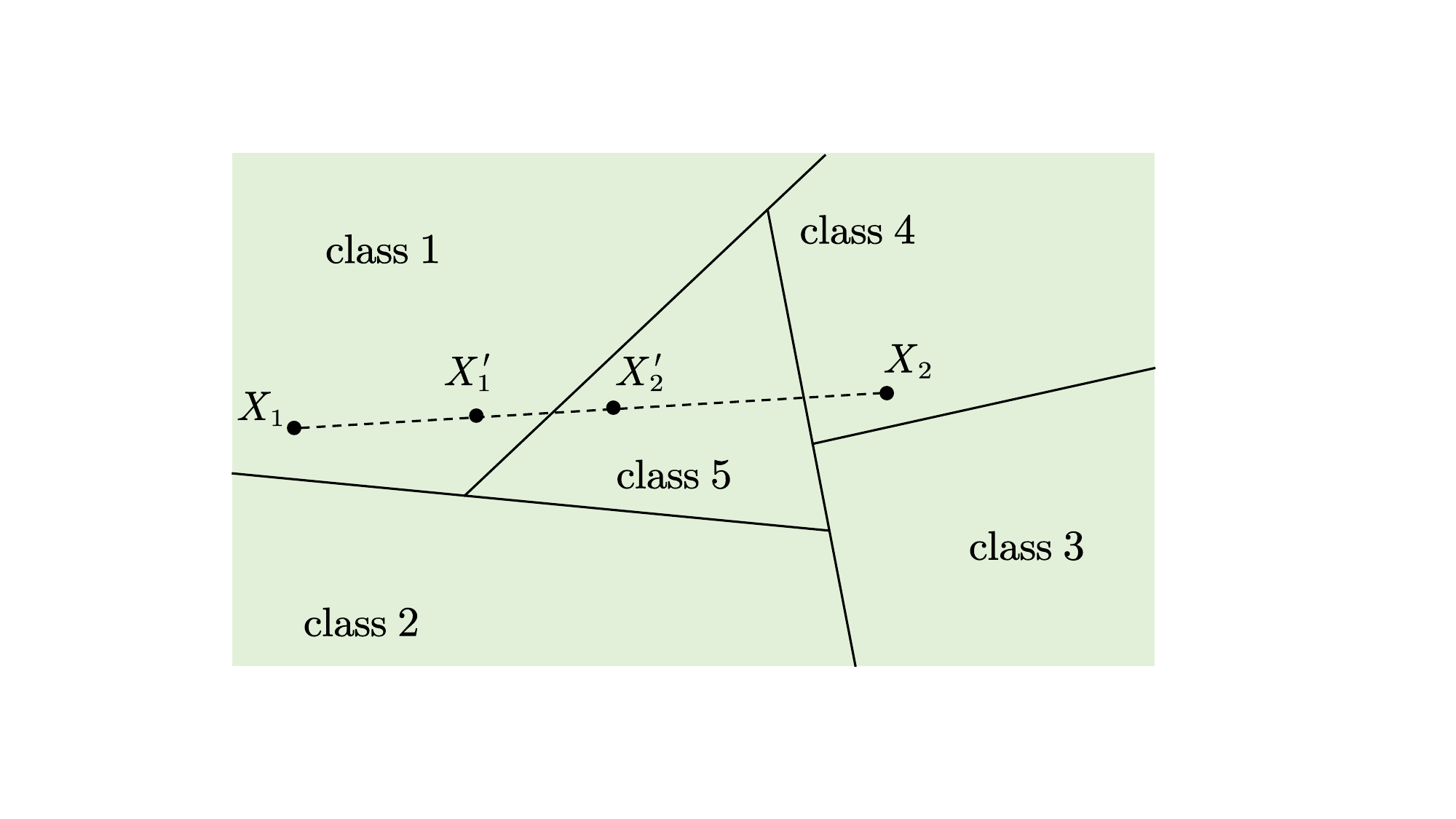}
\caption{
The idea of finding critical points 
when the feedback is top-$m$ scores.
}
\label{fig:improved_method_for_critical_points}
\end{figure}

Fig.~\ref{fig:improved_method_for_critical_points} 
shows an example in a 2-dimensional input space.
For $X_1$ (respectively, $X_2$) in the figure,
suppose that the top-$2$ labels are classes $1$ and $2$
(according to their distance from the regions of the two classes),
then we only know $y_2 - y_1$ for $X_1$.
For $X_2$, the top-$2$ labels are classes $3$ and $4$,
then we do not know $y_2 - y_1$ for $X_2$,
which prevents us from searching critical points based on $X_1$ and $X_2$.

As a result, we have to sample two random points again,
which wastes too many queries.
Moreover, we find that the above phenomenon occurs frequently
in model parameter recovery attacks.

\subsubsection{Finding Two Good Starting Points for Searching Critical Points. }
To solve the problem,
we propose improved binary search-based methods.
The core idea is to find two good starting points for searching critical points.
As depicted in Fig.~\ref{fig:improved_method_for_critical_points},
after randomly sampling two points $X_1$ and $X_2$,
let $X_1$ move along the direction $X_2 - X_1$,
and $X_2$ move along the opposite direction $X_1 - X_2$,
until they arrive at $X_1^{\prime}$ and $X_2^{\prime}$ respectively,
such that the number of elements of the intersection 
$\mathcal{L}_1 \cap \mathcal{L}_2$ exceeds $1$.
Then, let $X_1^{\prime}$ and $X_2^{\prime}$ be the two starting points.

This method significantly improves the success rate of finding critical points 
using the binary search-based method.  
However, this method can not prevent the increase in the query complexity,
and the model parameter recovery attack may still fail in some cases,
refer to the experiment results and analysis in the next section.



\section{End-to-End Practical Experiments}
\label{sec:practical_experiments}


The proposed model parameter recovery attacks
are evaluated by performing end-to-end
\footnote{
In~\cite{DBLP:conf/eurocrypt/CanalesMartinezCHRSS24, 
DBLP:conf/eurocrypt/CarliniCHRS25
}, 
the authors evaluate their attacks under an assumption 
that the first $i-1$ layers have been recovered \emph{without any precision errors}
when recovering the model parameters in layer $i$. 
The end-to-end attack stands for the attack executed without such an assumption, 
i.e., the recovery of layer $i$ must consider the errors of 
recovered parameters of the first $i-1$ layers.
The recovery of deeper layers is sensitive to 
the errors in prior layers~\cite{DBLP:conf/crypto/CarliniJM20,
DBLP:conf/asiacrypt/ChenDGSWW24,
DBLP:conf/nips/FoersterMSH24
}.
}
practical experiments
on diverse PReLU neural networks.
Concretely, following the training pipeline adopted 
by Carlini et al. in~\cite{DBLP:conf/crypto/CarliniJM20},
we first train several neural networks on random data,
and perform the end-to-end attack
on these neural networks.
Then, based on a typical benchmarking image dataset (i.e., MNIST)
in visual deep learning, we train several neural networks,
and perform the end-to-end attack on these models.
At last, we evaluate the proposed attacks in the case where
the feedback is (top-m) scores.

Note that the victim model $f_{\theta}$ is denoted by 
'$d_0$-$d_1$-$\cdots$-$d_{n+1}$', 
where $d_j$ is the dimension in layer $j$.

\subsection{Computing $(\varepsilon, 0)$-Functional Equivalence}
For any input $X \in \mathbb{R}^{d_0}$,
let $r = \mathrm{max} \{ \left| f_{\theta} (X) - f_{\widehat{\theta}} (X) \right| \}$
denote the maximum difference between the raw output of the victim model 
and that of the recovered model.
To quantify the degree to which a model parameter extraction attack has succeeded,
we adopt the concept of $(\varepsilon, \xi)$-functional 
equivalence~\cite{DBLP:conf/crypto/CarliniJM20}.

\begin{myDef} 
[\textbf{\textup{$(\varepsilon, \xi)$-Functional Equivalence}}~\cite{DBLP:conf/crypto/CarliniJM20}]
\label{def:functional_equivalence}
Two neural networks $f$ and $g$ are 
$(\varepsilon, \xi)$-functionally equivalent on $\mathcal{S}$ if
\begin{equation}
\mathrm{Pr}_{X \in \mathcal{S}} \left[ r \leqslant \varepsilon \right] \geqslant 1 - \xi.
\end{equation}
\end{myDef}

\subsubsection{Error Bounds Propagation. }
The method (i.e., error bounds propagation) proposed in~\cite{DBLP:conf/crypto/CarliniJM20}
is used to compute the $(\varepsilon, 0)$-functional equivalence of 
the recovered neural network $f_{\widehat{\theta}}$.
Its idea is to compare the recovered parameters 
$\widehat{A}^i$, $\widehat{B}^i$, $\widehat{S}^i$
with the real parameters
$A^i$, $B^i$, $S^i$
and derive an upper bound on the error of the raw output.

Since the recovered model is equivalent to the victim model,
the recovered parameters (except for the slopes) 
are not the same as the real parameters,
and the neuron order may also be changed.
Therefore, before comparing the model parameters,
we first align them.
Since all the technical details (e.g., how to align model parameters,
and propagate error bounds layer-by-layer)
are the same as that introduced in~\cite{DBLP:conf/crypto/CarliniJM20},
we do not introduce them again in this paper.

\subsection{Experiments on Neural Networks Trained on Random Data}
Table~\ref{tab:results_of_raw_output_attack} summarizes the experimental results
on PReLU neural networks trained on random data,
which verifies the effectiveness of the proposed attacks.

\begin{table}[!htbp]
\centering
\renewcommand\arraystretch{1.1}
\caption{Experiment results on raw output-based model parameter recovery attacks.}
\label{tab:results_of_raw_output_attack}
\begin{tabular}{c|c|c|c|c|c}
\hline
Architecture	&	Parameters		&	Workflow	&
Queries		&	$(\varepsilon, 0)$	&	max$|\theta - \widehat{\theta}|$	\\
\hline
32-16-1		&		561		&	1	&
$2^{15.61}$	&	$2^{-33.89}$	&	$2^{-36.93}$	\\
			&				&	2	&
$2^{15.74}$	&	$2^{-33.20}$	&	$2^{-36.15}$	\\
\hline
64-32-1		&		2145		&	1	&
$2^{16.94}$	&	$2^{-32.06}$	&	$2^{-35.92}$	\\
			&				& 	2	&
$2^{17.81}$	&	$2^{-31.63}$	&	$2^{-34.94}$	\\
\hline
128-64-1		&		8385		&	1	&
$2^{19.90}$	&	$2^{-29.28}$	&	$2^{-33.10}$	\\
			&				&	2	&
$2^{19.34}$	&	$2^{-29.55}$	&	$2^{-33.79}$	\\
\hline
20-10-10-1	&		351		&	1	&
$2^{18.19}$	&	$2^{-12.43}$	&	$2^{-15.36}$	\\
			&				&	2	&
$2^{15.51}$	&	$2^{-31.62}$	&	$2^{-33.44}$	\\
\hline
32-16-16-1	&		849		&	1	&
$2^{19.38}$	&	$2^{-0.85}$	&	$2^{-4.90}$	\\
			&				&	2	&
$2^{16.40}$	&	$2^{-19.79}$	&	$2^{-23.96}$	\\
\hline
64-32-32-1	&		3233		&	1	&
$2^{21.35}$	&	$2^{-5.20}$	&	$2^{-9.01}$	\\
			&				&	2	&
$2^{18.64}$	&	$2^{-23.14}$	&	$2^{-27.44}$	\\
\hline
\multicolumn{6}{l}{
max$|\theta - \widehat{\theta}|$: the maximum extraction error
of model parameters. 
}		\\
\end{tabular}
\end{table}

The results in Table~\ref{tab:results_of_raw_output_attack} 
also demonstrate the superiority of the neuron splitting technique-based method 
discussed in Section~\ref{subsec:joint_recovery}. 
When compared with Workflow $1$ 
(where the neuron sign and slope are recovered using the approach outlined in Section~\ref{subsec:independent_recovery}), 
Workflow $2$ (which employs the neuron splitting technique-based 
method described in Section~\ref{subsec:joint_recovery} 
for recovering the neuron sign and slope) significantly enhances the attack performance.

Table~\ref{tab:error_distributions} further shows the maximum error 
of the recovered slopes of each layer.
It is clear that the maximum recovery errors ($\max |S^i - \widehat{S}^i|$)
in the first workflow are much larger than those in the second workflow.

\begin{table}[!htbp]
\centering
\renewcommand\arraystretch{1.0}
\caption{Maximum error of the recovered slopes of each layer.}
\label{tab:error_distributions}
\begin{tabular}{c|c|c|c|c|c|c}
\hline
Architecture	&	\multicolumn{2}{c|}{20-10-10-1}			
			&	\multicolumn{2}{c|}{32-16-16-1}		
			&	\multicolumn{2}{c}{64-32-32-1}		\\
\hline
Workflow		&	1		&	2	&	1	&	2	&	1	&	2	\\
\hline
max$|S^1 - \widehat{S}^1|$	&	$2^{-24.48}$	&	$2^{-36.14}$	
						&	$2^{-18.96}$	&	$2^{-34.31}$	
						&	$2^{-20.56}$	&	$2^{-30.50}$		\\
\hline	
max$|S^2 - \widehat{S}^2|$	&	$2^{-15.75}$	&	$2^{-33.44}$	
						&	$2^{-4.90}$	&	$2^{-23.96}$	
						&	$2^{-9.01}$	&	$2^{-27.44}$		\\
\hline
\end{tabular}
\end{table}

\subsection{Experiments on Neural Networks Trained on MNIST}

MNIST is a typical benchmarking dataset used in visual deep learning.
It contains ten classes of 
handwriting number gray images~\cite{DBLP:journals/pieee/LeCunBBH98}. 
Each of the ten classes, i.e., `0', `1', `2', `3', `4', `5', `6', `7', `8', and `9',
contains $28 \times 28$ pixel gray images, 
totaling $60000$ training and $10000$ testing images.

When classifying different classes of objects,
the resulting neural networks will be different in terms of their
model parameters and mathematical properties, etc.
To fully verify the proposed attacks, 
we divide the ten classes into five groups
and build a binary classification neural network for each group.
We perform a standard rescaling of the pixel values from
$0 \cdots 255$ to $0 \cdots 1$.
Typical model training settings (the loss is the cross-entropy loss; 
the optimizer is standard stochastic gradient descent; 
batch size $128$) are adopted.
The first four columns of 
Table~\ref{tab:attack_results_on_mnist} summarize a detailed 
description of the neural networks to be attacked in this section.
We follow the idea in~\cite{DBLP:conf/crypto/CarliniJM20},
i.e., verifying the attack on diverse model architectures.
Therefore, Table~\ref{tab:attack_results_on_mnist} reports experimental results
on different model architectures.

\begin{table}[!htb]
\centering
\renewcommand\arraystretch{1.1}
\caption{Experiment results on neural networks 
trained on MNIST. }
\label{tab:attack_results_on_mnist}
\begin{tabular}{c|c|c|c|c|c|c}
\hline
task		&	architecture	&	accuracy		&	parameters	&
Queries		&	$(\varepsilon, 0)$	&	
max$|\theta - \widehat{\theta}|$	\\
\hline
`0' vs `1'		&	784-40-1		&	0.9990		&	31481		&
$2^{21.48}$	&	$2^{-31.61}$	&	$2^{-34.34}$	\\
\hline
`2' vs `3'		&	784-50-1		&	0.9765		&	39351		&
$2^{21.95}$	&	$2^{-27.42}$	&	$2^{-29.66}$	\\
\hline
`4' vs `5'		&	784-60-1		&	0.9952		&	47221		&
$2^{22.10}$	&	$2^{-29.34}$	&	$2^{-32.03}$ 	\\
\hline
`6' vs `7'		&	784-50-1		&	0.9965		&	39351		&
$2^{21.89}$	&	$2^{-18.06}$	&	$2^{-21.13}$ 	\\
\hline
`8' vs `9'		&	784-40-1		&	0.9778		&	31481		&
$2^{21.21}$	&	$2^{-31.61}$	&	$2^{-35.23}$ 	\\
\hline
`0' vs `1'		&	196-20-15-1	&	0.9991		&	4306			&
$2^{19.01}$	&	$2^{-26.65}$	&	$2^{-27.98}$	\\
\hline
`2' vs `3'		&	196-20-15-1	&	0.9672		&	4306			&
$2^{19.14}$	&	$2^{-23.77}$	&	$2^{-27.95}$	\\
\hline
`4' vs `5'		&	196-15-15-1	&	0.9936		&	3241			&
$2^{18.96}$	&	$2^{-25.54}$	&	$2^{-29.52}$ 	\\
\hline
`6' vs `7'		&	196-20-20-1	&	0.9945		&	4421			&
$2^{19.35}$	&	$2^{-25.57}$	&	$2^{-26.13}$ 	\\
\hline
`8' vs `9'		&	196-20-10-1	&	0.9722		&	4191			&
$2^{19.35}$	&	$2^{-11.14}$	&	$2^{-12.20}$ 	\\
\hline
\multicolumn{6}{l}{
max$|\theta - \widehat{\theta}|$: the maximum extraction error
of model parameters. 
}		\\
\multicolumn{6}{l}{
accuracy: classification accuracy of the victim model $f_{\theta}$. 
}		\\
\multicolumn{6}{l}{
196-$d_1$-$d_2$-$d_3$: the images are downsampled to a size of $14 \times 14$.
}		\\
\end{tabular}
\end{table}

In Table~\ref{tab:attack_results_on_mnist},
the last three columns summarize the experimental results,
which are obtained based on Workflow $2$,
i.e., the neuron signs and slopes are recovered by the neuron splitting technique-based 
method introduced in Section~\ref{subsec:joint_recovery}.

The experimental results show that the proposed attack
achieves good performance on these neural networks trained on the real gray images
in a realistic environment,
further verifying the effectiveness of the proposed attack.
The experimental results also show that the attack performance 
(i.e., the value of $\varepsilon$ and max$|\theta - \widehat{\theta}|$)
is influenced by the concrete model parameters.
We do not find any deeper quantitative relationship between the attack performance
and the model parameters.
Considering that the unknown relationships 
do not affect the verification of the proposed attack,
we leave it as future work.

There is an extra finding in experiments on 2-deep neural networks.
If we increase the input dimension to $784$,
the precision refinement algorithm in~\cite{DBLP:conf/crypto/CarliniJM20}
fails to reduce parameter errors for some neurons
(no identifiable common features observed),
preventing the recovery of deeper layers.
This phenomenon occurs frequently in large-scale PReLU neural networks.
However, this phenomenon seems not to be caused by PReLU activations,
since we also observe it in large-scale ReLU neural networks.
Readers can verify this phenomenon in large-scale ReLU neural networks
via the code provided in~\cite{DBLP:conf/crypto/CarliniJM20}.
This finding suggests a potentially interesting research direction for understanding fundamental limitations in attacking large-scale neural networks.


\subsection{Experiments on Scores-based Attack}
The last experiment is performed to support the argument that
the model parameters can be recovered 
by transforming the (top-$m$) scores-based attack scenario into one 
where the raw output is accessible. 
Table~\ref{tab:result_of_score_based_attacks} shows the experimental results
on three neural networks trained on random data and the MNIST dataset,
which supports the above argument.
Besides, the results in Table~\ref{tab:result_of_score_based_attacks} 
also prove that the proposed technique (i.e., finding good starting points) 
can improve the robustness and weaken the negative influence, 
since for each victim model in most cases (i.e., different settings of $m$), 
the performances of our attack are similar, 
i.e., not only recovering the model parameters 
but also requiring similar query complexity.


\begin{table}[!htbp]
\centering
\renewcommand\arraystretch{1.3}
\caption{Results of scores-based model parameter recovery attacks.}
\label{tab:result_of_score_based_attacks}
\begin{tabular}{c|c|c|c|c|c|c|c|c|c}
\hline
$m$			&	\multicolumn{3}{c|}{32-16-10}			
			&	\multicolumn{3}{c|}{20-10-10}	&	
\multicolumn{3}{c}{784-10-$10^{\star}$}\\
\cline{2-10}
			&	Queries	&	$(\epsilon, 0)$	&	max$|\theta - \widehat{\theta}|$	
			&	Queries	&	$(\epsilon, 0)$	&	max$|\theta - \widehat{\theta}|$
			&	Queries	&	$(\epsilon, 0)$	&	max$|\theta - \widehat{\theta}|$		\\
\hline
10			&	$2^{24.54}$	&	$2^{-31.91}$	&	$2^{-35.27}$	
			&	$2^{22.92}$	&	$2^{-33.31}$	&	$2^{-36.05}$
			&	$2^{19.77}$	&	$2^{-10.49}$	&	$2^{-15.96}$		\\
\hline
9			&	$2^{23.92}$	&	$2^{-31.75}$	&	$2^{-35.11}$	
			&	$2^{22.20}$	&	$2^{-34.45}$	&	$2^{-36.75}$
			&	$2^{19.56}$	&	$2^{-10.46}$	&	$2^{-15.90}$		\\
\hline
8			&	$2^{23.18}$	&	$2^{-33.01}$	&	$2^{-36.41}$	
			&	$2^{22.34}$	&	$2^{-33.42}$	&	$2^{-34.79}$
			&	$2^{19.95}$	&	$2^{-10.53}$	&	$2^{-17.52}$		\\
\hline
7			&	$2^{23.54}$	&	$2^{-32.02}$	&	$2^{-35.29}$	
			&	$2^{21.94}$	&	$2^{-33.04}$	&	$2^{-33.29}$
			&	$2^{19.44}$	&	$2^{-10.25}$	&	$2^{-14.05}$		\\
\hline
6			&	$2^{23.68}$	&	$2^{-31.80}$	&	$2^{-34.19}$	
			&	$2^{21.88}$	&	$2^{-33.98}$	&	$2^{-36.18}$
			&	$2^{19.71}$	&	$2^{-10.41}$	&	$2^{-16.30}$		\\
\hline
5			&	$2^{24.06}$	&	$2^{-31.92}$	&	$2^{-35.33}$	
			&	$2^{21.59}$	&	$2^{-32.69}$	&	$2^{-32.93}$
			&	$2^{19.64}$	&	$2^{-10.49}$	&	$2^{-15.62}$		\\
\hline
4			&	$2^{24.67}$	&	$2^{-30.82}$	&	$2^{-33.98}$	
			&	$2^{22.19}$	&	$2^{-32.14}$	&	$2^{-34.62}$
			&	$2^{20.33}$	&	$2^{-10.17}$	&	$2^{-15.26}$		\\
\hline
3			&	$2^{24.91}$	&	$2^{-15.80}$	&	$2^{-19.41}$	
			&	$2^{23.02}$	&	$2^{-32.49}$	&	$2^{-34.49}$
			&	$2^{20.45}$	&	$2^{-10.07}$	&	$2^{-17.00}$		\\
\hline
2			&	$2^{24.99}$	&	$2^{-31.06}$	&	$2^{-33.13}$	
			&	-			&	- 			&	-		
			&	$2^{21.27}$	&	$2^{-9.77}$	&	$2^{-14.07}$\\
\hline
\multicolumn{7}{l}{
-: the model parameter recovery attack failed.  
}		\\
\multicolumn{7}{l}{
$\star$: the model is trained on the MNIST dataset. 
}		\\
\end{tabular}
\end{table}

There are some unusual experimental results 
as presented in Table~\ref{tab:result_of_score_based_attacks}.
Consider the victim model $20$-$10$-$10$. 
When the feedback is top-$2$ scores, our attack failed. 
After tracing the attack process, 
we find that the attack is trapped in finding critical points to recover the slopes. 
In other words, the feedback makes some critical points inaccessible, 
which breaks the diversity of critical points and further leads to the attack failure.
Note that the victim model, instead of the feedback,
is the fundamental factor causing this unusual result.
Consider the victim model $32$-$16$-$10$.
When the feedback is top-$3$ scores, 
the precision of the recovered parameters is not as high as in other cases.
This result is related to the victim model.
At the same time, it is also related to the setting of $m$,
since $m$ influences the search for critical points,
which in turn affects the attack performance. 
In our attack, we randomly
choose $2$-out-of-$m$ scores each time to find critical points.
If we search for critical points by a fixed strategy 
(e.g., always using top-$2$ scores),
worse results (e.g., the attack on $20$-$10$-$10$ 
would fail regardless of $m$) may occur.

The above unusual experimental results show that
the negative influence of the scores-based feedback 
on the raw output-based attacks is complex 
and related to the properties of the concrete victim model. 
Considering that these unknown complex relations do not change the conclusion
that it is feasible to recover the model parameters of PReLU neural networks
under the feedback of (top-$m$) scores,
we leave the study of these unknown relations to future research.

\section{Conclusion}
\label{sec:conclusion}
This paper makes a step forward in the field of cryptanalytic extraction of neural networks.
For PReLU activation function-based fully connected neural networks,
we propose effective and efficient model parameter recovery attacks under three different
scenarios where the feedbacks are the raw output, all the probability scores,
and the top-$m$ probability scores, respectively.
End-to-end practical experiments on PReLU neural networks 
with a wide range of model architectures show the efficacy of the proposed attacks.
To the best of our knowledge,
this is the first time to prove with practical end-to-end evaluations that
it is possible to recover a neural network equivalent to a PReLU neural network,
only given access to the raw output, all the probability scores, even the top-$m$ scores.

Our work in this paper opens several directions for future research. 
First, developing attacks for highly expansive ReLU 
and PReLU neural networks is an important research topic.
Second, improving the proposed attack, 
especially improving the robustness of the attacks in the scenario 
where only the (top-$m$) scores are accessible, is an important direction.
Third, how to recover the model parameter of PReLU neural networks 
in the scenario where the feedback is the most likely class label is still unknown.
At last, whether it is possible to extract new neural networks 
with more complex model architectures is an interesting and tough question.

\subsubsection{Acknowledgments. }
We would like to thank the anonymous reviewers
from CRYPTO 2025 and ASIACRYPT 2025 for their detailed
and helpful comments.
This work was supported by the Key R\&D Program 
of Shandong Province (2024ZLGX05), 
the National Cryptologic Science Fund of China 
(2025NCSF02041, 2025NCSF02014), 
Tsinghua University Dushi Program, 
Zhongguancun Laboratory,
and Scientific Research Innovation Capability Support 
Project for Young Faculty (ZYGXQNJSKYCXNLZCXM-P4).
Yi Chen was also supported by the Shuimu Tsinghua Scholar Program.


\appendix



\section{Properties of PReLU Neural Networks}
\label{sec:properties_of_prelu}

\subsection{Equivalence of PReLU Neural Networks}

Permutation and scaling can help generate equivalent PReLU neural networks. 

\paragraph{$\mathbf{Permutation. }$}
The order of neurons in each layer of a network does not change
the underlying function. 
Formally, let $p_{k, \varrho} \left( f_{\theta} \right)$ be the network obtained from
$f_{\theta}$ by permuting layer $k$ according to $\varrho$.
This is equivalent to adjusting the order of columns (respectively, rows) of 
weights $A^k$ (respectively, $A^{k+1}$) according to $\varrho$ 
(along with the bias $B^k$ and slope $S^k$).
Then, $p_{k, \varrho} \left( f_{\theta} \right)$ is equivalent to $f_{\theta}$.

\paragraph{$\mathbf{Scaling. }$}
Scaling the incoming weights and biases of any neuron,
while inversely scaling the outgoing weights,
does not change the underlying function.
Formally, for the $i$-th neuron $\eta$ in layer $k$ and for any $c > 0$,
let $\mathcal{S}_{\eta, c} \left( f_{\theta} \right)$ be the network obtained from
$f_{\theta}$ by replacing $A_{i}^k$, $b_i^k$, and $A_{:, i}^{k+1}$
(i.e., the $i$-th row of $A^k$, the $i$-th element of $B^k$, 
and the $i$-th column of $A^{k+1}$)
by $c A_{i}^k$, $c b_i^k$, and $\frac{A_{:, i}^{k+1}}{c}$, respectively.
Then, $\mathcal{S}_{\eta, c} \left( f_{\theta} \right)$ is still equivalent to $f_{\theta}$,
refer to the following proof.

\begin{proof}
Let $\eta = \eta _1^k$ be the first neuron in layer $k$ without loss of generality.
For all $X \in \mathbb{R}^{d_0}$ and the neural network $f_{\theta}$, 
the preactivation input $y_j^{k+1}$ of the $j$-th neuron $\eta _j^{k+1}$ in layer $k+1$ is
\begin{small}
\begin{equation}	\label{eq:real_val}
\begin{aligned}
y_j^{k+1} &= a_{j, 1}^{k+1} \sigma_{k, 1} (y_1^k) + 
		    \sum_{i=2}^{d_k}{ a_{j, i}^{k+1} \sigma_{k, i} (y_i^k)} 	  \\
			&= a_{j, 1}^{k+1} \sigma_{k, 1} 
			      \left( 
			      \sum_{h=1}^{d_{k-1}}{ a_{1, h}^k \sigma_{k-1, h} (y_h^{k-1}) + b_1^k }  
			      \right) +
			      \sum_{i=2}^{d_k}{ a_{j, i}^{k+1} \sigma_{k, i} (y_i^k)}  .
\end{aligned}
\end{equation}
\end{small}
By comparison, for the neural network $\mathcal{S}_{\eta, c} \left( f_{\theta} \right)$,
we have
\begin{small}
\begin{equation}	\label{eq:new_val}
\begin{aligned}
y_j^{k+1} &= \frac{1}{c} a_{j, 1}^{k+1} \sigma_{k, 1} (y_1^k) + 
		    \sum_{i=2}^{d_k}{ a_{j, i}^{k+1} \sigma_{k, i} (y_i^k)} 	  \\
			&= \frac{1}{c} a_{j, 1}^{k+1} \sigma_{k, 1} 
			      \left( 
			      \sum_{h=1}^{d_{k-1}}{ c a_{1, h}^k \sigma_{k-1, h} (y_h^{k-1}) + c b_1^k }  
			      \right) +
			      \sum_{i=2}^{d_k}{ a_{j, i}^{k+1} \sigma_{k, i} (y_i^k)}   \\
			&= a_{j, 1}^{k+1} \sigma_{k, 1} 
			      \left( 
			      \sum_{h=1}^{d_{k-1}}{ a_{1, h}^k \sigma_{k-1, h} (y_h^{k-1}) + b_1^k }  
			      \right) +
			      \sum_{i=2}^{d_k}{ a_{j, i}^{k+1} \sigma_{k, i} (y_i^k)}  .
\end{aligned}
\end{equation}
\end{small}
where we used the property that $\sigma _{k, 1} (cx) = c \sigma_{k, 1} (x)$ for any $c > 0$.
Since the inputs $y_j^{k+1}$ in Eq.~\eqref{eq:real_val} and Eq.~\eqref{eq:new_val} are equal,
we conclude that $\mathcal{S}_{\eta, c} \left( f_{\theta} \right)$ 
is equivalent to $f_{\theta}$.
\end{proof}




\subsection{Linear Regions}
\label{subsec:linear_region}
The property of linear regions of PReLU neural networks contains two aspects.
First, the input space $\mathbb{R}^{d_0}$ can be partitioned into many linear regions.
Within each linear region $\mathcal{R}$,
for any neuron in the neural network,
any two inputs $X_1 \in \mathcal{R}$ and $X_2 \in \mathcal{R}$ 
make its neuron state the same.
Moreover, within a linear region,
each PReLU computes a fixed linear function
($\sigma _{k, i} (x) = x$ or $\sigma _{k, i} (x) = s_i^k \times x$),
then the whole PReLU neural network degenerates into an affine transformation.
Second, the hidden state space $\mathbb{R}^{d_k}$ 
(the input of the $k$-th activation layer)
can be partitioned into many linear regions.
Within each linear region, the PReLU neural network
degenerates into an affine transformation. 

%


%


\section{Auxiliary Algorithms}
\label{appendix:auxiliary_algos}

\subsection{Finding Critical Points and Precision Refinement}

To find critical points,
we adopt the binary search-based methods 
proposed in~\cite{
DBLP:conf/crypto/CarliniJM20,
DBLP:conf/eurocrypt/CanalesMartinezCHRSS24},
since these methods only rely on the property of linear regions.
After recovering the weights and biases,
the general method proposed in~\cite{DBLP:conf/crypto/CarliniJM20} 
is applied to refine precision of recovered weights and biases.

\subsection{Filtering Critical Points in Wrong Layers}
Consider the recovery of layer $i$.
After collecting numerous critical points,
a question is how to filter critical points not belonging to layer $i$.

To filter critical points belonging to layers $1, \cdots, i-1$,
we check the outputs of these layers directly,
since the parameters of these layers have been recovered.

To filter critical points belonging to deeper layers 
(i.e., layers $i+1, \cdots, n$), 
we count the frequency of weight vectors 
(i.e., $\widetilde{A}_j^i$) recovered by our attack, 
and only keep the $d_i$ weight vectors with the highest frequency.
Our attack is expected to return a random weight vector 
for critical points belonging to deeper layers. 
On the contrary, for critical points 
corresponding to the same neuron in layer $i$,
our attack returns the weight vector $\widetilde{A}_j^i$.
This is based on the truth that our attack is designed for layer $i$.
Detailed proof is presented in Appendix~\ref{appendix:filter_wrong_critical_points}.

%

\section{Filtering Critical Points in Deeper Layers}
\label{appendix:filter_wrong_critical_points}

Consider the raw output-based method proposed in
Section~\ref{subsec:recover_hidden_layer}
for recovering the weight of layer $i$.
Without loss of generality, 
suppose that $X = [x_1, \cdots, x_{d_0}]^{\top}$ is a critical point corresponding to
the first neuron in layer $i^{\prime} \in \{i+1, \cdots, n\}$. 

When we run the weight recovery method over this critical point, we have
\begin{equation}	
\delta = \left\{  \begin{array}{l}
g_{1}^{i^{\prime}+1} \left( 1 - s_1^{i^{\prime}} \right) A_1^{i^{\prime}}  H^{\prime},
\,\, \mathrm{if} \,\, A_1^{i^{\prime}}  H^{\prime} >0 ,\\
g_{1}^{i^{\prime}+1} \left( s_1^{i^{\prime}} - 1 \right) A_1^{i^{\prime}} H^{\prime} ,
\,\, \mathrm{if} \,\, A_1^{i^{\prime}}  H^{\prime} >0 .\\
\end{array}
\right.
\end{equation}
where $H^{\prime}$ is the perturbation direction in layer $i^{\prime} - 1$.\

The relation between $H^{\prime}$ and $H$ (the perturbation direction in layer $i - 1$) is
$H^{\prime} = A^{\prime} H + B^{\prime} $
where $A^{\prime}$ and $B^{\prime}$ are determined by
the $\sum_{j=i}^{i^{\prime} - 1}{d_j}$ 
neuron states in layers $i, i+1, \cdots, i^{\prime} - 1$.

According to the rationale of the method 
proposed in Section~\ref{subsec:recover_hidden_layer},
if randomly sampling enough perturbation $H_j^{\prime}$ in layer $i^{\prime} - 1$,
we can recover a weight vector $\widetilde{A}_1^{\prime}$ 
(i.e., the weight vector $A_1^{i^{\prime}}$ multiplied by a constant) 
by building and solving a system of linear equations,
such that $\widetilde{A}_1^{\prime} H_j^{\prime} = \alpha _j$.

However, we are trying to recover the weights of neurons in layer $i$,
therefore $H_j^{\prime}$ is replaced by the perturbation direction $H_j$ 
in layer $i$ when building a system of linear equations.
Since $H_j^{\prime} = A^{\prime} H_j + B^{\prime}$,
the weight recovery method in Section~\ref{subsec:recover_hidden_layer} 
will return a weight vector determined by $A^{\prime}$ and $B^{\prime}$,
i.e., determined by the $\sum_{j=i}^{i^{\prime} - 1}{d_j}$ 
neuron states in layers $i, i+1, \cdots, i^{\prime} - 1$.

Denote by $\mathcal{P}$ the set of all the neuron states
in layers $i, i+1, \cdots, i^{\prime} - 1$.
According to Lemma 2 
presented in~\cite{DBLP:conf/asiacrypt/ChenDGSWW24},
the number of possible $\mathcal{P}$'s
is close to $2^{\sum_{j=i}^{i^{\prime} - 1}{d_j}}$.
For the critical point $X$ belonging to layer $i^{\prime}$,
our weight recovery method is equivalent to choosing an element 
(i.e., a weight vector)
from a set consisting of about $2^{\sum_{j=i}^{i^{\prime} - 1}{d_j}}$ elements.
Apparently, when $2^{\sum_{j=i}^{i^{\prime} - 1}{d_j}}$ is larger,
it is less likely to choose the same elements in two trials.

Based on the above analysis, we have two conclusions.
First, given two critical points corresponding to different layers,
it is impossible to recover the same weight vectors.
Second, when we are recovering the weights of layer $i$,
given two critical points belonging to deeper layers,
even if they correspond to the same neuron,
the probability that our attack returns the same weight vectors
is also extremely low, or even approaching $0$.
Due to the two conclusions, 
in the proposed model parameter recovery attacks,
we filter critical points in deeper layers by counting the frequency
of weight vectors recovered by our attack.

%

\section{Attack Complexity of the Proposed Attacks}
\label{appendix:attack_complexity}

\subsection{Complexity of Raw Output-based Attack}
The complete model parameter recovery attack is performed layer by layer.
The recovery of each layer contains three steps: finding critical points,
recovering weight vectors, and recovering neuron signs and slopes (excluding the last layer).

Consider layer $i$.
First, 
we collect $\mathcal{O}\left( d_i \rm{log}_2 {d_i} \right)$ critical points
\footnote{According to the coupon collector's argument (assuming uniformity),
we will have at least two critical points corresponding to every neuron in layer $i$,
refer to the analysis in~\cite{DBLP:conf/crypto/CarliniJM20}.}.
After filtering critical points belonging to layers $1, \cdots, i-1$,
we recover a weight vector based on each critical point,
using the differential attack in Section~\ref{sec:signature_recovery_with_raw_output}.
Then, by counting the frequency of recovered weight vectors, 
we identify the correct weight vectors for the $d_i$ neurons.
At last, we recover the neuron sign and slope based on the weight vectors,
using the method in Section~\ref{subsec:joint_recovery}.

To find a critical point, we query the victim model $\epsilon$ times, 
where $\epsilon$ is a small constant determined by the binary search method 
in~\cite{DBLP:conf/crypto/CarliniJM20, DBLP:conf/eurocrypt/CanalesMartinezCHRSS24}.
To recover a weight vector based on a critical point,
we query the victim model $4 d_{i-1} -1$ times,
refer to the differential attack Section~\ref{sec:signature_recovery_with_raw_output}.
The recovery of neuron sign and slope does not involve any queries.
Therefore, the query complexity of the recovery of layer $i$ is
$\mathcal{O}\left( d_i (4 d_{i-1} - 1)\rm{log}_2 {d_i} \right)$.
As a result, the query complexity of the complete attack is also \emph{polynomial}.

The time complexity of the complete attack is also \emph{polynomial}.
Consider the recovery of layer $i$ again.
During the process of collecting critical points, 
querying the victim model almost occupies all the running time.
As for the weight recovery,
we query the victim model $4 d_{i-1} - 1$ times and solve a system of linear equations.
Neural networks contain numerous floating-point number operations,
therefore, the main time consumption also comes from querying the victim model.
Since the query complexity is polynomial,
the time complexity is also \emph{polynomial}.

\subsection{Complexity of Scores-based Attack}
Since the (top-$m$) scores-based attack is based on the raw output-based attack,
and the value of $m$ has no significant influence on the complexity,
the query and time complexity of the (top-$m$) 
scores-based attack are also \emph{polynomial}.

%
%
%
\bibliographystyle{splncs04}
\bibliography{reference}

\begin{thebibliography}{10}
\providecommand{\url}[1]{\texttt{#1}}
\providecommand{\urlprefix}{URL }
\providecommand{\doi}[1]{https://doi.org/#1}

\bibitem{DBLP:journals/tnn/Baum91}
Baum, E.B.: Neural net algorithms that learn in polynomial time from examples
  and queries. {IEEE} Trans. Neural Networks  \textbf{2}(1),  5--19 (1991)

\bibitem{DBLP:conf/eurocrypt/CanalesMartinezCHRSS24}
{Canales Martinez}, I.A., Ch{\'{a}}vez{-}Saab, J., Hambitzer, A.,
  Rodr{\'{\i}}guez{-}Henr{\'{\i}}quez, F., Satpute, N., Shamir, A.: Polynomial
  time cryptanalytic extraction of neural network models. In: Joye, M.,
  Leander, G. (eds.) EUROCRYPT 2024. Lecture Notes in Computer Science, vol.
  14653, pp. 3--33. Springer (2024)

\bibitem{DBLP:conf/eurocrypt/CarliniCHRS25}
Carlini, N., Ch{\'{a}}vez{-}Saab, J., Hambitzer, A.,
  Rodr{\'{\i}}guez{-}Henr{\'{\i}}quez, F., Shamir, A.: Polynomial time
  cryptanalytic extraction of deep neural networks in the hard-label setting.
  In: Fehr, S., Fouque, P. (eds.) EUROCRYPT 2025. LNCS, vol. 15601, pp.
  364--396. Springer (2025)

\bibitem{DBLP:conf/crypto/CarliniJM20}
Carlini, N., Jagielski, M., Mironov, I.: Cryptanalytic extraction of neural
  network models. In: Micciancio, D., Ristenpart, T. (eds.) CRYPTO 2020. LNCS,
  vol. 12172, pp. 189--218. Springer

\bibitem{DBLP:conf/asiacrypt/ChenDGSWW24}
Chen, Y., Dong, X., Guo, J., Shen, Y., Wang, A., Wang, X.: Hard-label
  cryptanalytic extraction of neural network models. In: Chung, K., Sasaki, Y.
  (eds.) ASIACRYPT 2024. LNCS, vol. 15491, pp. 207--236. Springer (2024)

\bibitem{DBLP:conf/iclr/DanielyG23}
Daniely, A., Granot, E.: An exact poly-time membership-queries algorithm for
  extracting a three-layer relu network. In: The Eleventh International
  Conference on Learning Representations, {ICLR} 2023, Kigali, Rwanda, May 1-5,
  2023. OpenReview.net (2023)

\bibitem{DBLP:journals/ijon/DubeySC22}
Dubey, S.R., Singh, S.K., Chaudhuri, B.B.: Activation functions in deep
  learning: {A} comprehensive survey and benchmark. Neurocomputing
  \textbf{503},  92--108 (2022)

\bibitem{DBLP:conf/nips/FoersterMSH24}
Foerster, H., Mullins, R.D., Shumailov, I., Hayes, J.: Beyond slow signs in
  high-fidelity model extraction. In: Globersons, A., Mackey, L., Belgrave, D.,
  Fan, A., Paquet, U., Tomczak, J.M., Zhang, C. (eds.) NeurIPS 2024 (2024)

\bibitem{DBLP:conf/iccv/HeZRS15}
He, K., Zhang, X., Ren, S., Sun, J.: Delving deep into rectifiers: Surpassing
  human-level performance on imagenet classification. In: ICCV 2015. pp.
  1026--1034. {IEEE} Computer Society (2015)

\bibitem{DBLP:conf/uss/JagielskiCBKP20}
Jagielski, M., Carlini, N., Berthelot, D., Kurakin, A., Papernot, N.: High
  accuracy and high fidelity extraction of neural networks. In: Capkun, S.,
  Roesner, F. (eds.) USENIX Security 2020. pp. 1345--1362. {USENIX} Association

\bibitem{DBLP:journals/pieee/LeCunBBH98}
LeCun, Y., Bottou, L., Bengio, Y., Haffner, P.: Gradient-based learning applied
  to document recognition. Proc. {IEEE}  \textbf{86}(11),  2278--2324 (1998)

\bibitem{DBLP:conf/icml/MartinelliSGB24}
Martinelli, F., Simsek, B., Gerstner, W., Brea, J.: Expand-and-cluster:
  Parameter recovery of neural networks. In: ICML 2024. OpenReview.net (2024)

\bibitem{DBLP:conf/fat/MilliSDH19}
Milli, S., Schmidt, L., Dragan, A.D., Hardt, M.: Model reconstruction from
  model explanations. In: Boyd, D., Morgenstern, J.H. (eds.) FAT* 2019.
  pp.~1--9. {ACM} (2019)

\bibitem{DBLP:conf/wpes/Reith0T19}
Reith, R.N., Schneider, T., Tkachenko, O.: Efficiently stealing your machine
  learning models. In: Cavallaro, L., Kinder, J., Domingo{-}Ferrer, J. (eds.)
  Proceedings of the 18th {ACM} Workshop on Privacy in the Electronic Society,
  WPES@CCS 2019, London, UK, November 11, 2019. pp. 198--210. {ACM} (2019)

\bibitem{DBLP:conf/icml/RolnickK20}
Rolnick, D., Kording, K.P.: Reverse-engineering deep relu networks. In: ICML
  2020. Proceedings of Machine Learning Research, vol.~119, pp. 8178--8187.
  {PMLR}

\bibitem{DBLP:conf/uss/TramerZJRR16}
Tram{\`{e}}r, F., Zhang, F., Juels, A., Reiter, M.K., Ristenpart, T.: Stealing
  machine learning models via prediction apis. In: Holz, T., Savage, S. (eds.)
  USENIX Security 2016. pp. 601--618. {USENIX} Association

\end{thebibliography}

\end{document}